\documentclass[letterpaper,twocolumn,10pt]{article}
\usepackage[T2A]{fontenc}
\usepackage{usenix}

\usepackage{array}
\usepackage{graphicx}
\usepackage{booktabs} 
\usepackage{listings}
\usepackage{xcolor}

\hypersetup{hidelinks}
\usepackage{amsmath}
\usepackage{cleveref}
\usepackage{xurl}
\usepackage{seqsplit}
\usepackage{tabularx}

\usepackage[table]{xcolor}
\usepackage{caption}
\usepackage{subcaption}

\usepackage{makecell}
\usepackage[russian,english]{babel}

\usepackage{longtable}
\usepackage[most]{tcolorbox}

\definecolor{codebg}{HTML}{F3F2FF}
\definecolor{codebar}{HTML}{9B95FF}

\lstdefinestyle{jsstyle}{
  basicstyle=\ttfamily\scriptsize,
  breaklines=true,
  columns=fullflexible,
  keepspaces=true,
  showstringspaces=false,
  frame=none,
  xleftmargin=0pt,
  xrightmargin=0pt,
  aboveskip=0pt,
  belowskip=0pt
}

\newtcblisting{maxcode}{
  listing only,
  listing options={style=jsstyle},
  colback=codebg,
  colframe=codebg,
  boxrule=0pt,
  arc=3pt,
  left=1.2em,
  right=1em,
  top=0.8em,
  bottom=0.8em,
  borderline west={2pt}{0pt}{codebar},
  enhanced,
  breakable
}

\usepackage{hyphenat}
\usepackage{fancyhdr}
\newcommand{\myparagraph}[1]{\vspace{5pt}\textbf{#1}\hspace{5pt}}
\usepackage{tikz}
\usepackage{amssymb}
\newcommand{\keyinsight}[1]{%
  \medskip\noindent
  \hangindent=1.2em\hangafter=1%
  {$\blacktriangleright$}\hspace{0.4em}{\textit{\textbf{Takeaway:}}} \textit{#1}%
  \smallskip
}

\begin{document}
%-------------------------------------------------------------------------------

%don't want date printed
\date{}

\title{Don't Trust the Super-App: A Case Study of Russia’s Max}

\newcommand{\umich}{\hspace{0.05em}$^1$}
\newcommand{\calgary}{\hspace{0.05em}$^2$}
\newcommand{\gatech}{\hspace{0.05em}$^3$}
\newcommand{\iitd}{\hspace{0.05em}$^4$}

\author{\parbox{\textwidth}{\centering\normalfont
Richa Priyanka\umich,
Aaron Ortwein\umich,
Joel Reardon\calgary,
Michael Specter\gatech,
Piyush Kumar Sharma\iitd,
Roya Ensafi\umich
\\[6pt]
\small
\umich \textit{University of Michigan}\quad
\calgary \textit{University of Calgary}\quad
\gatech \textit{Georgia Institute of Technology}\quad
\iitd \textit{Indian Institute of Technology, Delhi}\quad
}}

\maketitle

%-------------------------------------------------------------------------------
\begin{abstract}
Super-apps, an emerging mobile architecture, host third-party mini-apps inside a single app, allowing users to access diverse services. A decade of security research on the super-app ecosystem has all assumed super-apps to be a trusted intermediary. We argue this implicit trust is difficult to justify: China's WeChat is already shown to passively track its user's activity across mini-apps at extraordinary scale;  Russia's MAX's parent company is reported to be deeply entangled with the
state prosecution of online speech; and Iran's Bale was reported to be functioning in the world's longest internet shutdown due to its state-backed support. 

In this paper, we show how malicious super-apps have undeniable capabilities to silently undermine the security and privacy of mini-apps and users without leaving any trace. Using MAX as an example, we show how it can capture mini-app UI, read and write mini-app local storage, inject arbitrary JavaScript into a mini-app's runtime, mediate mini-app network traffic, and control authentication context in ways that can enable silent user impersonation. Sadly, these capabilities manifest themselves in any super-app because of the architectural privileges granted to them by design. We argue that mobile OS and app store interventions are urgently needed to close this architectural blind spot before it is further exploited.
\end{abstract}

\section{Introduction}

Super-apps, as popularized by WeChat, are an emerging mobile architecture in which a host app embeds third-party mini-apps, delivering diverse services and functionality within a single unified app~\cite{vandervlist2025superappification}.
Security researchers have scrutinized this model, demonstrating that current super-apps cannot consistently isolate mini-apps, enforce least-privilege access to sensitive resources, prevent cross-platform exploits, or contain data exposure at scale~\cite{zhang2021measurement, zhangL2022identity, zhangX2023overcollection, zhangJ2023smallleak, zhangY2023appsecret, wang2023hidden, lu2020resource, wang2025wechatknows, wang2023crossplatform, wang2024rootfree}. They also collaborated directly with super-app developers through responsible disclosure, resulting in
vendor patches, bug bounties, and fixes with
the goal of improving the super-app itself~\cite{zhangL2022identity,lu2020resource,wang2025wechatknows,wang2023crossplatform}.
While extremely valuable, prior work largely accepts the super-app as an OS layer atop the mobile OS, and assumes it can be trusted to act in the best interests of mini-app developers and users.
\begin{figure}[!t]
  \centering
  \includegraphics[width=0.8\columnwidth]{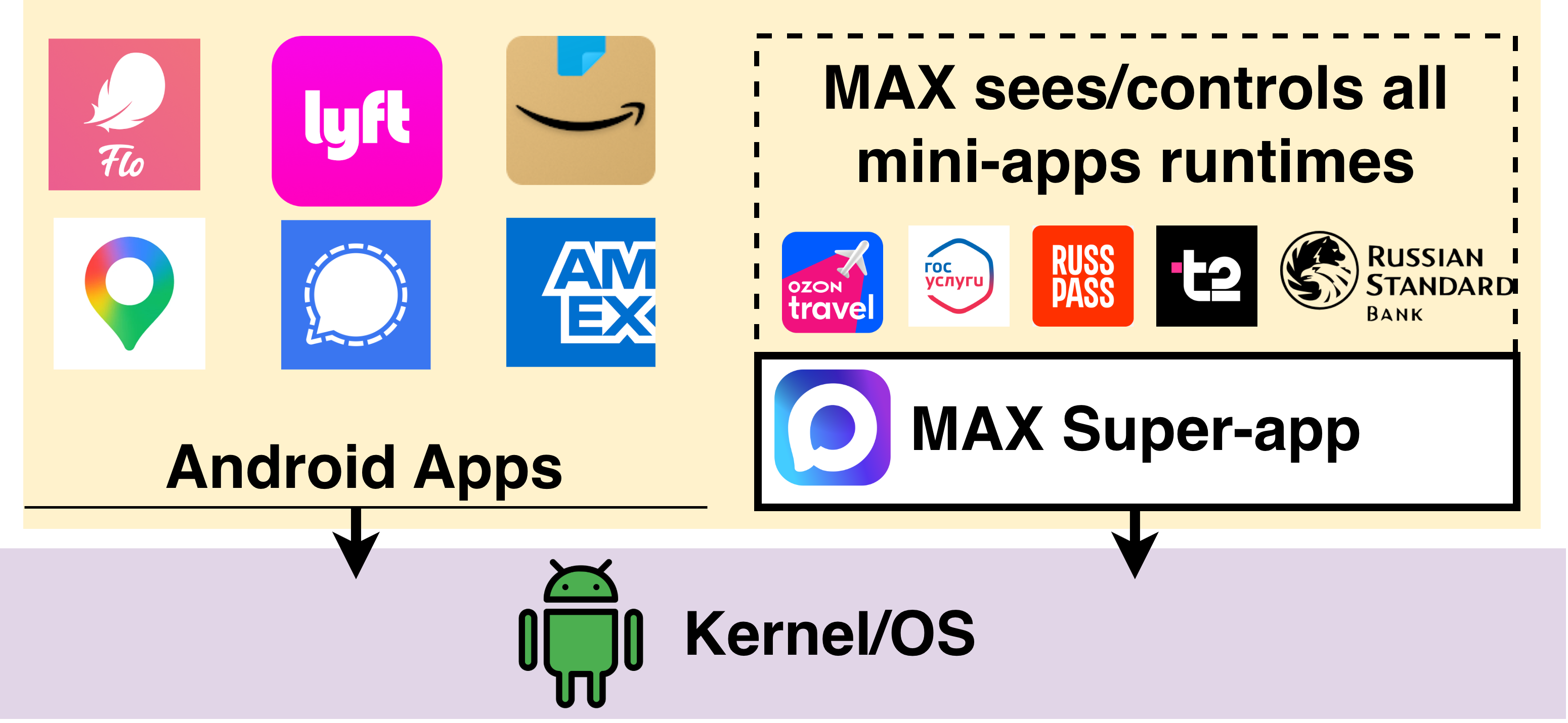}
  \caption{\textbf{Super-app Position in Android.} Android treats super-apps like normal apps, despite their privileged visibility into and control over hosted mini-apps.}
  \label{fig:super-app-arch}
\end{figure}
Treating super-apps as trusted intermediaries is dangerous as it ignores the global reality where information control is not just a feature of authoritarian regimes, but also a growing part of digital governance. In the West, the United Kingdom, Canada, and Sweden have recently pursued legislative measures that would, in practice, weaken or bypass end-to-end encryption on major platforms~\cite{washpost2025apple,eff2025apple,infosec2025signal,gec2025sweden,globemail2026signal,citizenlab2026billc22}.
Where platforms already operate under strong government pressure, a super-app can be an unusually efficient instrument for surveillance and censorship. Notably, Wang et al. demonstrated that WeChat \textit{passively} tracks user activity across its mini-app ecosystem at extraordinary scale~\cite{wang2025wechatknows}, and reports from Iran (Bale), Vietnam (Zalo), and India (Paytm) document parallel trajectories in which government-aligned platforms have become deeply embedded in everyday communication and commerce~\cite{otf2024iran,tuoitre2025zalo,orf2018paytm,buzzfeed2018paytm}.
\begin{figure*}[t]
  \centering
  \includegraphics[width=.75\textwidth]{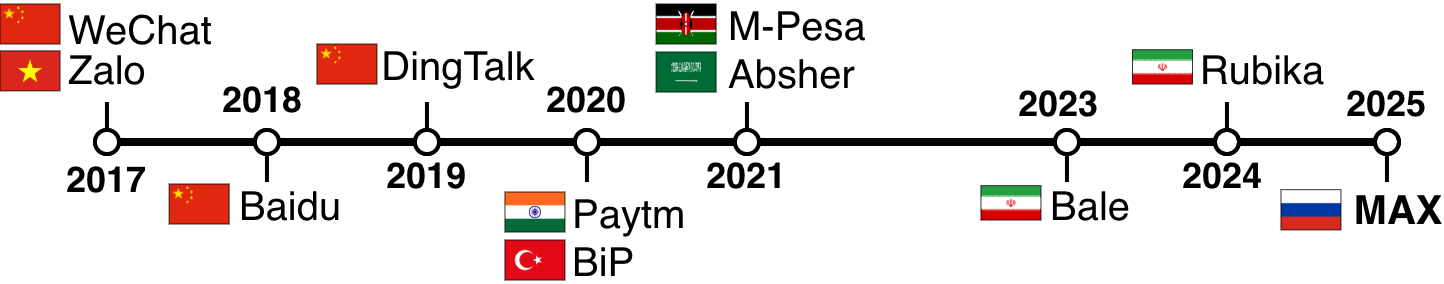}
\caption{\textbf{Launch Timeline of Selected Super-apps.} Sources: WeChat~\cite{knockel2020wechat}, Zalo~\cite{tuoitre2025zalo}, Baidu~\cite{scmp2018,citizenlab2023}, DingTalk~\cite{freedom2021,alibaba2023}, Paytm~\cite{orf2018paytm,shinde2021}, BiP~\cite{turkishminute2021}, M-Pesa~\cite{reuters2025}, Absher~\cite{hrw2019}, Bale~\cite{otf2024iran}, Rubika~\cite{otf2024iran}, and MAX~\cite{cnn2025max}.}
\label{fig:superapp-timeline}
\end{figure*}
These trends make the implicit trust placed in super-apps increasingly difficult to justify, yet no prior work has systematically examined what happens when that trust is \textit{actively} violated by the super-app itself. Under government pressure, how can a super-app \textit{actively} undermine the security of its users and mini-apps? Without a user's knowledge, can it interfere with their mini-app activities in real time? Without a mini-app's knowledge, can it forge connections on behalf of its users? And, most importantly, does the super-app hand authorities the client-side backdoor they have long sought to obtain through conventional legislative action?

\begin{figure*}[t]
  \centering
  \includegraphics[width=\textwidth]{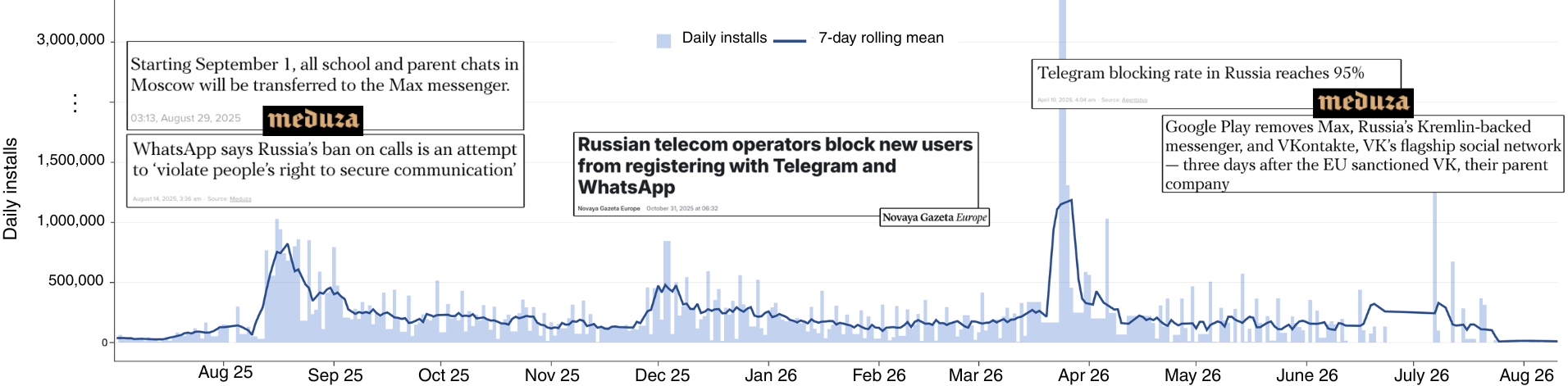}   
  \caption{\textbf{Daily Google Play Store Installs for the MAX Super-app (June 2025--June 2026).}}
  \label{fig:max-avg-installs}
\end{figure*}

In this paper, we aim to show how a malicious super-app can abuse its privileged position to design \textit{active} attacks that undermine the confidentiality, integrity, and availability of information. Specifically, we illustrate this threat through an in-depth analysis of MAX, Russia's state-backed super-app that has been coercively forced on Russian citizens since March 2025. MAX is developed by VK whose flagship social network, VKontakte, is deeply entangled with the state prosecution of online speech.
Journalists have documented that over a fourteen-year period, VK was involved in nearly 46\% of criminal cases related to social media posts in Russia and is now sanctioned by the European Union over abusive surveillance~\cite{dietrich2023vk,novayagazeta2024,euvksanction26}. Worse, MAX is no longer optional in Russia, as it has become the sole gateway to access essential public services like school, payments, and digital identity. We focus on MAX not just because of the above reasons, but also because Russia's rapid descent towards digital authoritarianism showcases a blueprint any government can follow, making MAX both the most urgent and the most exportable case for studying the super-app architecture.

However, conducting such an investigation is not straightforward as MAX employs several mechanisms that impede systematic analysis. MAX offers a different APK in Russia (pre-installed on Russian phones) compared to what is available via the Google Play Store, and a large set of mini-apps only function in Russian networks. We found MAX performs location detection by cross-referencing VPN API signals, SIM card metadata, GPS coordinates, and other device identifiers, making it difficult to execute the app in a controlled analysis environment without triggering detection. Moreover, instead of transmitting data in standardized formats such as JSON or XML, MAX employs a custom RPC protocol for its network communication. Therefore even after the traffic is decrypted, standard parsing techniques fail to yield intelligible output, necessitating a full reverse-engineering effort. To overcome these challenges, we designed and implemented a custom end-to-end analysis pipeline to instrument the app, capture all events, and reconstruct MAX's runtime behavior from a Russian proxy and a local US network.

Playing the role of an active adversary, we found five distinct capabilities that allow MAX to act as a man-in-the-middle for all mini-app interactions: (1) MAX can capture screenshots of mini-app content without holding any special system permissions, and without alerting the user; (2) It holds full read and write access to all mini-app local storage, meaning no data a mini-app persists on-device is private from MAX; (3) It injects JavaScript into running mini-apps, enabling silent, undetectable modification of mini-app functions at run-time; (4) It mediates all mini-app network traffic, and in the Russian regional build specifically, routes this traffic through a GOST TLS proxy, raising acute concerns about state-level interception~\cite{rfc9189}; (5) Finally, it controls the authentication tokens and session context supplied to each mini-app, granting it the ability to silently impersonate any user to any service hosted within the super-app ecosystem. These findings uncover that MAX's super-app architecture can actively and silently undermine the security guarantees that users assume when interacting with each mini-app.

Unfortunately, as we show in Section \ref{sec:androidvssuperapps}, these capabilities do not require exploitation of a vulnerability but are a consequence of the \textit{architectural privileges} granted to super-apps by design. Consequently, these capabilities are not specific to MAX but are inherent to any implementation of the super-app architecture. Android's security model treats the super-app and all its mini-apps as a single, unified trust principal, providing no mechanism to constrain, audit, or observe the super-app exercising these capabilities. The consequence is that hundreds of millions of users interact with what appear to be independent, sand-boxed services, while in reality operating entirely within a single app.

Any progress on advancing super-app security has to first take into account that super-apps are an active threat. We therefore propose recommendations aimed for the mobile OS and app store in Section \ref{sec:recommendations}. In the near term, the OS should provide users controls to revoke network access and background execution. In the longer term, OS constraints should be developed to enable better accountability, so super-app capabilities match the security and privacy assumptions of mini-apps and users. To this end, we recommend adding user-facing transparency mechanisms, auditability over super-app to mini-app communication, and architectural isolation, so that mini-apps can establish trust independently of the super-app.

\section{Background}

Super-apps are mobile applications in which a single native app provides the runtime environment, permission mediation layer, and distribution channel for third-party services, commonly called mini-apps. Super-apps allow users to conveniently access diverse services through one app, often with a shared account. This model has been adopted widely by WeChat, Alipay, Telegram, TikTok, and others \cite{alipaytechcrunch,tiktok,telegramreuters,zhang2021measurement}

The proliferation of this architecture reflects convergent incentives for both
the super-app providers and mini-app developers. Mini-apps can be developed
with standard web technologies (HTML, JavaScript, CSS), distributed instantly to
the super-app's existing user base without per-release app store review, and
gain access to native device capabilities such as cameras, contacts, and
location through the super-app's bridge APIs. These mini-apps also increase user
engagement and retention on the super-apps. 

A user who pays bills, messages with
friends, and books transit on the same app is less likely to abandon the app than one using standalone apps for each service. The super-app can
also profit from mini-apps by charging fees to all payment transactions that occur through the super-app's payment infrastructure.
Finally, daily use of the super-app also sustains its privileges because keeping it in foreground preserves permissions Android would otherwise restrict to foreground-only or auto-revoke after disuse.

The same consolidation that makes super-apps commercially successful also makes them appealing to authorities. This is why state-aligned super-app models are emerging in different countries (Figure~\ref{fig:superapp-timeline}). In Iran, government-backed apps such as Rubika and Bale reportedly lack meaningful end-to-end encryption, and route user-clicked URLs through backend servers for monitoring~\cite{otf2024iran}. In Vietnam, Zalo serves 79 million monthly users under expanded data collection terms, legally bound to comply with government data requests under the Cyber Security Law~\cite{tuoitre2025zalo}. Similarly, in India, Paytm was reported to have shared user data from the geopolitically sensitive Jammu and Kashmir region with the Prime Minister's office, operating under a legacy legal framework that provides minimal safeguards for electronic data~\cite{orf2018paytm,buzzfeed2018paytm}.
Because super-apps concentrate payments, communication, government services, and large facets of daily life, disabling a user's account becomes a way to enforce partial expulsion from society, as seen from the effects of Tencent blocking WeChat accounts in 2022~\cite{wechatmit2022}.

MAX makes this concern more pressing. Launched by VK in March 2025, MAX messenger was positioned to absorb functions that extend beyond private communication to include payments, school and government services~\cite{cnn2025max}. Figure~\ref{fig:max-avg-installs} shows that MAX's adoption rose in distinct waves that align with Russia's disruption of competing messengers. The first major surge occurs in mid-August 2025, when the 7-day rolling mean rises to roughly 823,000 installs per day. This coincides with Roskomnadzor, Russia’s
communication agency, restricting voice and video calls on WhatsApp and Telegram, after which MAX briefly became the top-ranked app in Russia~\cite{meduzawhatsapp2025, iz2025maxtop}. Around this time, all school chats also started being transferred to MAX~\cite{meduzaschool2025}. A second smaller wave appeared in late November and early December 2025, when the rolling mean reached roughly 480,000 installs per day. This followed the widespread Telegram disruptions and widening restrictions on foreign communication apps like FaceTime~\cite{novaya2025registration}. The largest spike appears in late March 2026, when daily installs peaked near 2.9 million amid major Telegram outages and reports of state employees, and public institutions being pushed onto MAX~\cite{meduzatelegramblock2026, mt2026telegram, france24_2026max}. Together, this pattern shows how MAX's growth has been influenced by a sequence of events where the Russian government degraded services for all competitive messengers while promoting a preferred domestic super-app.

These developments have also prompted substantial scrutiny of MAX among independent security researchers and digital-rights groups. Since its launch, researchers have inspected MAX's Android permissions and compared them with other messengers, examined its APK and embedded third-party libraries, and performed network measurements to characterize the services and endpoints it contacts~\cite{chumikov_max_permissions_2025,chumikov_max_network_2025,grishutin_max_apk_2025,cyberb_max_apk_2025}. Other community-led analyses have examined telemetry and environment-detection logic exposed through static reverse engineering, while RKS Global conducted controlled tests of MAX's access to sensitive device resources and later analyzed its VPN-detection behavior~\cite{3ntr0py_max_audit_2026,rks_max_surveillance_2025,rks_russian_apps_vpn_2026}. These independent investigations reflect the considerable suspicion surrounding MAX's rapid expansion and have helped document aspects of its permissions, bundled components, and observable network and application behavior.

\subsection{Super-app Architecture}

Beyond the standard components of a mobile application, a super-app introduces three architectural elements that are central to its operation: the host application, the mini-app layer, and the JavaScript bridge (Figure ~\ref{fig:super-app-arch}).

\myparagraph{The Host Application}
The host application is a native mobile app that is installed from a platform app store (Google Play, Apple App Store, RUStore, Huawei App Gallery, etc.). It provides a WebView-based execution environment for the mini-apps it hosts, a JavaScript bridge layer exposing native capabilities, a lifecycle manager, a mini-app marketplace and launcher, and a permission mediation layer governing mini-app access to device resources. In addition to managing the runtime, the host typically provides shared platform services like authentication, payment processing, and push notifications that multiple mini-app developers can invoke through encapsulated APIs. This abstraction is a key part of the model's appeal: a mini-app developer calls a single bridge method (e.g. \texttt{requestPayment()}) rather than integrating with platform specific SDKs, substantially reducing development effort and time to deployment.
\begin{figure}[!t]
  \centering
  \includegraphics[width=\columnwidth]{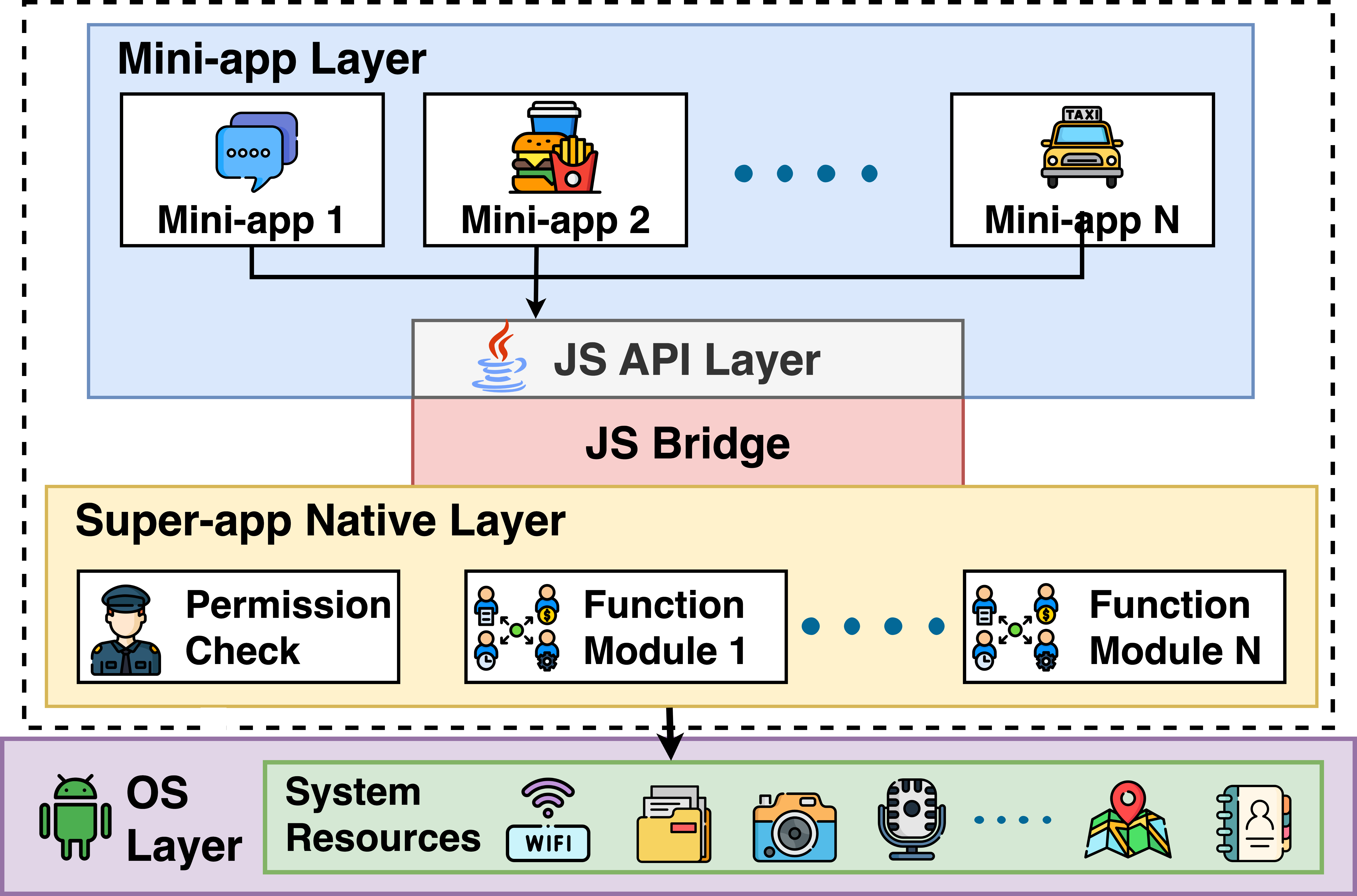}
  \caption{\textbf{Super-app Architecture.} Mini-apps call a shared JavaScript API Layer, which crosses a JavaScript Bridge to the Native Layer for permission checks and access to OS-level system resources.}
  \label{fig:super-app-arch}
\end{figure}

\myparagraph{The Mini-app Layer}
Mini-apps are light-weight applications that execute within the host's runtime environment. Unlike native mobile applications, mini-apps are built with web-based technologies (HTML, CSS, JavaScript). They are rendered inside a WebView, which is a browser component provided by the platform OS (Android's \texttt{WebView}, iOS's \texttt{WKWebView}) that lets native components render and execute web content within their process~\cite{webview}. The host instantiates and configures the WebView, making it the execution sandbox for all hosted mini-apps.

Mini-apps do not require any OS level installation and can be discovered and launched from the host's marketplace. The host app manages the fetching, caching and entire lifecycle of the mini-app. The mini-app also does not have its own application package, UID, or a dedicated storage directory \cite{androidsecurityplatform}. It operates entirely within the host's OS-level identity and sandbox. Access to native device capabilities like camera, contacts, location, storage, is mediated by the host through its bridge layer rather than granted by the OS.

\myparagraph{The JavaScript Bridge}
The JavaScript bridge is the central architectural element of the super-app model. It defines the API surface available to mini-apps and mediates all interaction between web-based mini-app code and native device capabilities. It is the mechanism by which mini-app code running in a WebView invokes native capabilities provided by the host. Since the WebView executes standard web content, it has no direct access to device hardware or OS services. The bridge provides this access by exposing host-implemented methods to the mini-app's JavaScript context.

On Android, the host application makes calls to \texttt{addJavascriptInterface()} to bind a Java object into the WebView, making its methods callable from JavaScript. A mini-app invokes these methods to access native functionality like reading from device storage or requesting a contact, and the host executes the corresponding native operation. This bridge is bidirectional: in the reverse direction, the host can call \texttt{evaluateJavascript()} to execute arbitrary JavaScript within the mini-app's WebView context. This allows the host to inject scripts, modify page behavior, or invoke callbacks.

\subsection{Prior Work}
A growing body of security and privacy research has substantially advanced our understanding of super-apps. At the ecosystem level, measurement work has characterized the scale and structure of these apps, showing how large mini-app marketplaces have grown and how much functionality super-apps expose to third-party developers~\cite{zhang2021measurement}. This shows the breadth of the attack surface a single super-app inherits when it mediates millions of mini-apps.

A second line of work has shown that this architecture is difficult to secure. Studies have demonstrated identity confusion, hidden or undocumented APIs, permission bypasses, unauthorized access to sensitive resources, privilege escalation, and UI deception attacks, each reproduced across billion-user apps \cite{zhangL2022identity,wang2023hidden,zhangJ2023smallleak,lu2020resource}. Cross-platform analyses further show that the same super-app can expose different security properties across mobile and desktop clients, leading to several possible exploits \cite{wang2023crossplatform,wang2024rootfree}. More recently, malware studies have examined the supply side of these ecosystems, showing that malicious mini-apps can evade super-app vetting and affect users at scale \cite{yang2025malware}. Finally, Wang et al. analyze super-app design through the lens of browser security, focusing primarily on isolation among mini-apps within a shared host runtime \cite{wangBetterSuperAppArchitecture2023}. 

Closest to our work is a growing line of privacy research that turns from malicious mini-apps to the privileged position of the super-app host itself. These studies show that mini-app ecosystems enable leaked secrets and systematic privacy over-collection \cite{zhangY2023appsecret,zhangX2023overcollection}. Recent work on first-party tracking in WeChat and inference attacks over mini-app interaction histories shows that the super-app can passively observe user behavior across all its mini-apps \cite{wang2025wechatknows,cai2025secrets}.

Our work builds on existing academic research on super-apps, but adopts a different threat model. Prior work largely treats the super-app as a flawed but trustworthy intermediary between the mini-apps and the OS. In that model, the adversary is typically a malicious mini-app, an external attacker, or a bug to be patched. Even the work that explores a super-app collecting a wide array of private data, studies them as a passive observer \cite{wang2025wechatknows}.

In this paper, we aim to ask what happens when the host itself becomes adversarial. Rather than asking only what the host can observe, we study whether it can actively act as a man-in-the-middle for all mini-app interactions in real-time without any safeguards.

\section{Methodology}
\label{sec:methodology}

\newcommand{\code}[1]{\texttt{\seqsplit{#1}}}

\begin{figure*}[t]
  \centering
  \includegraphics[width=\textwidth]{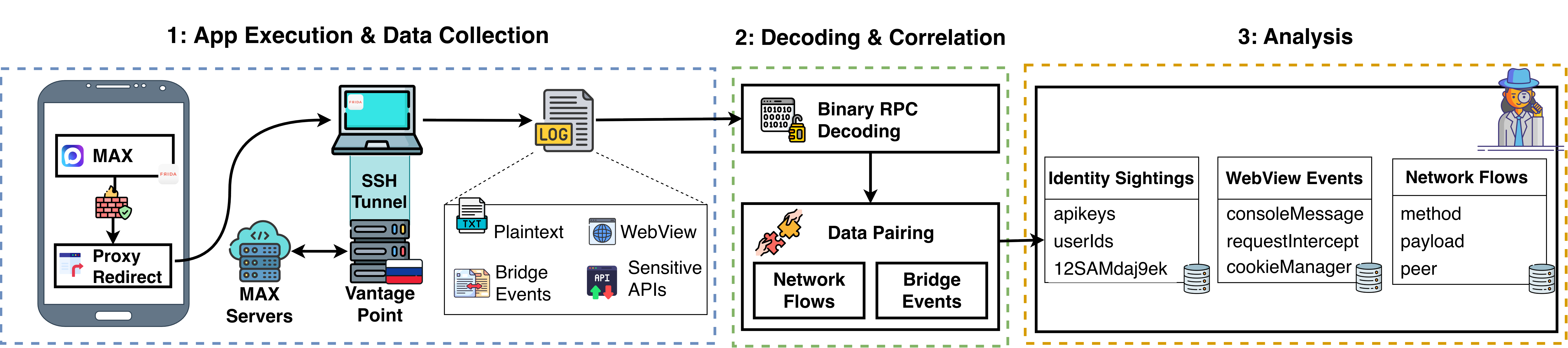}
  \caption{\textbf{Measurement Pipeline.} In stage 1, we instrument MAX on an Android device with Frida and tunnel its traffic to a controlled vantage point while collecting plaintext logs. In stage 2, we decode binary RPC messages, and stitch network and bridge events together. In stage 3, we store the synthesized results like WebView events and network flows into a SQLite database for further analysis.}
  \label{fig:methodology}
\end{figure*}

In this section, we present our setup and the detailed methodology for analyzing the MAX super-app. Figure~\ref{fig:methodology} shows our high-level analysis pipeline for setting up a test environment, executing the MAX app, collecting raw data, processing it, and then performing systematic analysis. We now explain in detail each part of our analysis.

\myparagraph{Threat Model}
Our adversary is a super-app host, that either because of malicious intent or under coercion, wants to abuse its privileged position. The adversary's goal is to compromise the confidentiality, integrity, and availability of mini-app interactions. We explore the capabilities that this adversary has by design. We assume that the mini-apps installed on this super-app are benign, and want to provide legitimate services to their users via the super-app. The victim uses the super-app normally and has at least one mini-app installed. 

\subsection{Test Environment}
\label{subsec:method-test-environment}

Our test environment consists of OnePlus mobile devices, rooted and configured with appropriate instrumentation for analysis (\Cref{subsec:data-collection-instrumentation}).
We obtain the  MAX APK through two channels: our Russian collaborators (henceforth called the Russian build) and the Google Play Store (henceforth called the universal build). After static analysis, we learned that these APKs were mostly similar, except for the cryptographic layer.

In our initial testing, the universal build operated without any obvious issues or connection restrictions. In contrast, the Russian build, once launched, remained indefinitely stuck at the login screen. On further investigation, we found that the backend servers (\texttt{api-gost.oneme.ru} on ports 443 and 8443, serving authentication, configuration, messaging, and media channels) were unreachable from non-Russian IPs. We confirmed this by observing that direct connections from our measurement machine were timing out. Without a Russian vantage point, the app only loaded a login screen but did not allow us to authenticate, rendering it unusable.

The Russian build restricts access through location checks, including client IP, device location, VPN interfaces, timezone, and SIM/network identifiers. Hence, we carefully designed an egress tunnel to an endpoint in Russia that operated outside of Android's VPN feature.
MAX queries \code{NetworkCapabilities.hasTransport(TRANSPORT\_VPN)} during session setup, and so we used \texttt{iptables NAT REDIRECT} to transparently route MAX's traffic through our proxy. In all subsequent captures from this build, MAX’s VPN probe returned \texttt{false}. To address other signals that reveal our real location, we configured the device timezone to \texttt{Europe/Moscow}, locale to \texttt{ru\_RU}, and emptied the SIM slot. We also cleared the data directory for MAX to avoid any persistent identifiers that could link sessions. After these steps MAX loaded successfully.

Finally, activating an account on MAX requires SMS verification. Earlier versions of MAX (until 25.14) restricted this registration to Russian phone numbers, which made account creation impractical for our test environment because new Russian SIMs require identity verification \cite{moscowtimes2025}. We registered test accounts and began dynamic testing after version 25.15 (released November 2025), which expanded registration to the Commonwealth of Independent States.

\subsection{Instrumentation}
\label{subsec:data-collection-instrumentation}
The challenge in instrumenting a super-app is that the app mediates mini-app activity across multiple layers including TLS-encrypted RPC to its own backend, JavaScript bridge calls between the super-app and mini-apps, callbacks that fire from the mini-app's WebViews, and Android system APIs that access sensitive resources. To get a complete picture of the app's runtime behavior, we designed a unified capture script that attaches to the running MAX process via Frida and arms hooks across four functional layers. These hooks emit logs prefixed with a tag to indicate the type of event logged. At start-up, the script detects the app build by probing for \texttt{ru.CryptoPro.ssl.SSLSocketImpl}. If this class is present, the script enables the Russia-specific GOST TLS hooks. Otherwise, it falls back to BoringSSL hooks.

The implementation had to address two sources of brittleness. First, arming 80+ hooks simultaneously destabilizes Android's just-in-time (JIT) compiler. The process sometimes crashes during garbage collection because Frida’s runtime patching can leave invalid metadata for JIT-compiled frames. To avoid this, we disable optimized execution for boot-image methods at script startup, forcing them through the interpreter. This trades some performance for more reliable instrumentation. Second, MAX’s R8/ProGuard obfuscation changes short class and method names across releases, so hooks tied to obfuscated symbols fail on update. Rather than bind hooks only to obfuscated symbols, we identify targets using stable string-constant fingerprints in decompiled smali, such as locating the bridge dispatcher by the method body containing \texttt{WebAppDeviceStorage*} constants. We then maintain a per-version symbol table, with class enumeration as a fallback when these symbols change.

\myparagraph{Plaintext TLS Capture}
\label{subsubsec:tls}
MAX's primary communication channel is a persistent TLS connection carrying a custom binary RPC protocol. We attach hooks to MAX’s TLS read/write routines, at the point where plaintext enters or leaves the cryptographic stack. This captures inbound payloads after decryption and outbound payloads before encryption. Since the two regional builds use different TLS stacks, the plaintext capture is build-specific. 

In the Russian build, MAX uses CryptoPro's pure-Java implementation of GOST TLS, which uses Russia's own national cryptographic algorithms and certificate authorities \cite{rfc9189}. We hook \texttt{SSLSocketImpl.read} and \texttt{write}, which sit at the terminal I/O boundary of the CryptoPro stream. At this point, inbound bytes have already been decrypted after GOST cipher processing, and outbound bytes have not yet been encrypted. Since CryptoPro's abstraction hides which server each TLS session connects to, we hook the socket-creation path to recover this mapping. This design is necessary because the Russian build’s GOST TLS stack is incompatible with standard interception tooling like mitmproxy, Burp Suite, and Wireshark’s \texttt{SSLKEYLOGFILE} workflow, which do not natively support the required GOST cipher suites. Additionally, a proxy-based approach would also require a trusted GOST-capable certificate chain. Hooking MAX’s own TLS API surface is therefore the only practical way to observe Russian build's plaintext.

The universal build follows Android's conventional Conscrypt/BoringSSL path. There, we hook native \texttt{SSL\_read} and \texttt{SSL\_write} in the device's \texttt{libssl.so}, capturing plaintext below the Java \texttt{SSLSocket} abstraction and immediately after Conscrypt's JNI bridge. To join each decrypted payload with endpoint metadata, we add companion hooks on \texttt{connect} and \texttt{getpeername} and reconstruct the mapping from SSL session to socket and remote peer. 

\myparagraph{Bridge Surface Enumeration}
In our capture script, we hook WebView functions \texttt{addJavascriptInterface} and \texttt{removeJavascriptInterface} to detect JavaScript bridge registration and teardown. When a bridge is registered, the hook reflects over the Java object and enumerates all methods annotated with \texttt{@android.webkit.JavascriptInterface}. Each registration emits a bridge enumeration log with the interface name, the associated obfuscated class, full signatures of JavaScript-callable methods, and exposed-method counts. This gives us the full bridge surface as the app declares it, including any methods we would have missed during static analysis alone.

To capture mini-app to MAX calls, we hook the bridge method \texttt{postEvent} on each registered bridge class. We also hook MAX's internal dispatcher which receives every \texttt{postEvent} invocation regardless of which bridge class it entered through. Each dispatch emits a bridge dispatch log containing the verb (named operation strings routing to a specific plugin handler), JSON payload, originating WebView URL, public bridge flag, and any application-level correlation identifier such as \texttt{requestId} or \texttt{queryId}.

To capture MAX to mini-app events, we hook the WebView function \texttt{evaluateJavascript}, which is used by MAX to inject code into the mini-app context. This records bridge responses, including host-injected \texttt{WebApp.sendEvent} calls, as well as host-installed code such as the \texttt{navigator.share} polyfill and FCP telemetry probes. Each injection emits a JS injection log with the injected source and target URL. This also gives us redundant coverage of the bridge. Across all sessions, every per-bridge \texttt{postEvent} record matches a dispatcher record with the same payload, and every dispatcher record has a corresponding bridge-entry record.

\myparagraph{WebView Lifecycle}
We hook MAX's WebView callback surface to observe how it mediates mini-app navigation, resource loading, trust decisions, and browser UI\@. We capture \texttt{shouldInterceptRequest} on resource requests from mini-apps, logging the URL, method, and headers with rate limiting to avoid excessive output. We capture \texttt{shouldOverrideUrlLoading} on page navigations, and \texttt{onPageStated}/\texttt{onPageFinished} to mark page-loads. For error and certificate handling, we hook \texttt{onReceivedHttpError}, \texttt{onReceivedError}, \texttt{onReceivedSslError}, and \texttt{onReceivedClientCertRequest}, recording HTTP failures, network errors, TLS validation failures and host responses, and mutual-authentication challenges.

We also capture MAX's mediation of browser features exposed to mini-apps.  When a mini-app tries to open a popup window, MAX intercepts the request and redirects it to the system browser. We hook this interception point to log where popups are sent. Hooks on \texttt{onJsAlert}, \texttt{onJsConfirm}, and \texttt{onJsPrompt} record JavaScript dialogs. Hooks on \texttt{onShowFileChooser} record mini-app file-upload requests, including accepted MIME types, capture flags, and selection mode. The mini-app WebView's console output is captured through \texttt{onConsoleMessage}.

Finally, we instrument WebView storage access from MAX's side. \texttt{CookieManager} hooks capture Java-intiated cookie reads, writes, removals, and flushes, while \texttt{WebStorage} hooks capture attempts to enumerate, query, or delete origin-scoped DOM storage. To catch direct access that bypasses these APIs, we also hook \texttt{FileInputStream}, \texttt{FileOutputStream}, and \texttt{RandomAccessFile} constructors and filter for paths under \texttt{app\_webview} in MAX's data directory, including cookies, IndexedDB, local storage and service worker state. The same file-output hooks emit logs for likely download destinations, allowing us to trace whether \texttt{WebAppDownloadFile} writes bytes into MAX's private storage or user-visible download paths.

\myparagraph{Sensitive APIs}
To document MAX’s access to privacy-sensitive Android resources, we hook system APIs across several categories: clipboard access, contact lookups, telephony and carrier metadata, location access, camera and microphone use, network-state and VPN-detection checks, package enumeration, and file I/O\@. For each observed access, the script records the API category, method invoked, arguments, return value, and a truncated stack trace. 

\subsection{App Execution \& Data Collection}

Each measurement session consisted of a single Frida-attached run of MAX\@. We connected the device to the lab laptop over USB, launched MAX, attached our instrumentation script, and then used the app normally while the script captured both the foreground and background activity from the MAX process. Each session produced a plaintext capture file containing event logs and session metadata such as app version, timestamp, test location, and network configuration.

Within a given session, we searched for mini-apps from inside MAX, opened the corresponding chatbot, launched the mini-app through its visible ``Open App'' icon, interacted with it, and then closed it before moving to the next one. For each mini-app, we performed a set of standardized interactions: opening the bot chat, launching the mini-app WebView, waiting for the page to load, and exploring the primary user-visible flows available without external identity verification. Where supported, we performed browsing, search, form-entry, and navigation actions for each mini-app. While we interacted with the mini-app, the Frida script captured MAX activity across layers, including sensitive API access, JavaScript injections into mini-app contexts, WebView navigation and lifecycle callbacks, bridge dispatches, and network traffic.

In each session we interacted with around 8--10 mini-apps. This kept individual logs small enough for downstream parsing and manual validation while still allowing us to observe multiple mini-app launches with the same device, account, build, and network conditions. Unlike other well-known super-apps, MAX has no documented public catalog of available mini-apps. Therefore, we iteratively searched mini-apps on the MAX app manually.  Altogether, we exercised 90 mini-apps, spanning healthcare, government services, banking and finance, telecom, commerce, education, and tourism (refer to Table \ref{tab:max-miniapp-bots} in Appendix for the full list of mini-apps we tested).

\subsection{Decoding and Correlation}
\label{subsec:decoding-and-correlation}
 
This section describes the steps we take to convert the raw text logs captured during data collection stage into structured, cross-referenced events suitable for analysis.

\myparagraph{Binary RPC Protocol}
Instead of using HTTP for the API traffic, MAX maintains a persistent TLS connection (\texttt{api-gost.oneme.ru} in the Russian build, \texttt{api.max.ru} in the universal build) and exchanges data using a custom binary RPC protocol. This is a purpose-built binary wire format that standard MITM proxies cannot parse without custom protocol dissectors. We reverse-engineered the protocol and discovered that each message consists of a 10-byte header followed by a MessagePack-encoded body (see Table \ref{tab:max-rpc-header} for details). 

\begingroup
\rowcolors{2}{gray!10}{white}
\begin{table}[!t]
\centering
\scriptsize
\setlength{\tabcolsep}{4pt}
\renewcommand{\arraystretch}{1.15}

\begin{tabularx}{\columnwidth}{
  @{}
  r
  >{\ttfamily}l
  >{\raggedright\arraybackslash}X
  @{}
}
\toprule
\rowcolor{white}% 
\textbf{Offset} & \textbf{Field} & \textbf{Description}\\
\midrule
0  & magic        & Always \texttt{0x0a} (protocol version 10).\\
1  & direction    & \texttt{0x00} request, \texttt{0x01} response.\\
2  & request\_id  & 2-byte big-endian ID to match responses to requests.\\
4  & opcode       & 2-byte big-endian RPC method identifier, e.g., \texttt{0x0001} heartbeat, \texttt{0x0013} init-sync, \texttt{0x0023} contacts presence, \texttt{0x0080} send-message.\\
6  & flags        & \texttt{0}: MessagePack body; non-zero: 2-byte vendor prefix precedes MessagePack.\\
7  & body\_length & 3-byte big-endian payload length, excluding the header.\\
10 & body         & MessagePack payload; if \texttt{flags} $\neq$ \texttt{0}, MessagePack starts after the 2-byte vendor prefix.\\
\bottomrule
\end{tabularx}
\caption{\textbf{Max Binary RPC Frame Header Layout.}}
\label{tab:max-rpc-header}
\end{table}
\endgroup

\myparagraph{TLS Session Correlation}
Our capture script emits each TLS read/write event as a structured triplet. \texttt{[EVT]} records metadata such as TLS/SSL session ID, sequence number, direction, byte count, peer/local addresses, and stream class. \texttt{[DATA]} carries the base64-encoded plaintext payload, prefixed with a 16-byte fingerprint for fast filtering. \texttt{[PROBE]} provides an ASCII preview for quick inspection before offline parsing. The analysis pipeline joins these records by \texttt{(ssl\_id, seq, dir)} and decodes the binary RPC envelope from the captured payload. In practice, this layer captures MAX's TLS traffic to its API endpoints, along with other host-originated encrypted flows such as TracerSDK crash telemetry, VK analytics pings, and CDN fetches. By design, we are not aiming to directly hook in to the WebView process to observe the raw mini-app traffic directly, since that is something MAX would not be able to do either.

\myparagraph{Bridge Request--Response Pairing}
The bridge protocol does not explicitly indicate whether a
\texttt{WebApp.sendEvent()} call is responding to a mini-app request or
delivering an event initiated by MAX\@. Both are delivered through the same
\texttt{evaluateJavascript} path, so request responses and host-initiated events
appear structurally similar in the trace. To distinguish them, our decoder pairs
bridge dispatch requests with JS injected responses using application-level
identifiers such as \texttt{requestId}, and emits bridge pair records with
round-trip latency. Unmatched injections are retained as standalone
host-to-mini-app events for later analysis.

\subsection{Analysis}

Finally, in the analysis stage, we ingest the decoded and stitched events from the previous stage (Section \ref{subsec:decoding-and-correlation}) into a SQLite database with tables to group similar events together for easier analysis. Events table contains all events logged. Network flows contain network events materialized from decoded and stitched TLS events. Identity Sightings keep track of authentication related parameters captured across all events, like user ID, API keys, etc. WebView Events keeps track of WebView callbacks across all mini-apps exercised during instrumentation. Every time the analysis script parses a new plaintext capture file, it also stores important context flags for account, region, device, app version, and build.

This database turns raw hook output into a cross-session view of MAX’s behavior. It lets us connect events that are otherwise scattered across logs: which endpoints were contacted, which RPC methods were invoked, which identifiers appeared in which flows, which Android APIs were accessed, and what WebView activity occurred around the same time. This is what allows us to make claims about data flows, identifier persistence, sensitive-resource access, and regional/build differences rather than relying on isolated log excerpts.

\section{Results}
\label{tab:results}

We now discuss MAX's potential capabilities based on our analysis database that contains a cross-session view of MAX's behaviors. Each subsection focuses in detail on an architectural capability and the security threat it enables. We first outline the default Android architecture that supports the underlying capability. Then we zoom into MAX's specific implementation and show how the security threat can be exploited.

As discussed in Section \ref{subsubsec:tls}, MAX ships two concurrent builds that are mostly identical but differ at the cryptographic layer. Both builds implement the same RPC protocol and mini-app mediation design, including the bridge API, JavaScript injection path, storage and network mediation. Since super-app security issues stem from this common design, we present our results from the Russian build only for simplicity. 

\subsection{Control Over Mini-app UI}

In Android, the mini-app WebView is a \texttt{View} embedded into the super-app view surface and its rendered output is ultimately presented inside a window owned by the super-app process. Although the WebView executes in an isolated process, this does not make the UI opaque to the super-app because it still owns the \texttt{View} and receives the final visual output.

Given this architectural capability, we ask if MAX can take a screenshot of its mini-app UIs. To check, we tried two screen capture paths from within MAX’s process using Frida. First, we tested Canvas-based capture by invoking \texttt{View.draw(Canvas)} on the mini-app UI from within MAX's process~\cite{androiddraws}. This method was not consistently reliable, and it only captured screenshots 73\% of the times due to limitations of software View for hardware-accelerated WebViews~\cite{softdrawdeprecate}.

\begin{figure}[!t]
  \centering
  \includegraphics[width=\columnwidth]{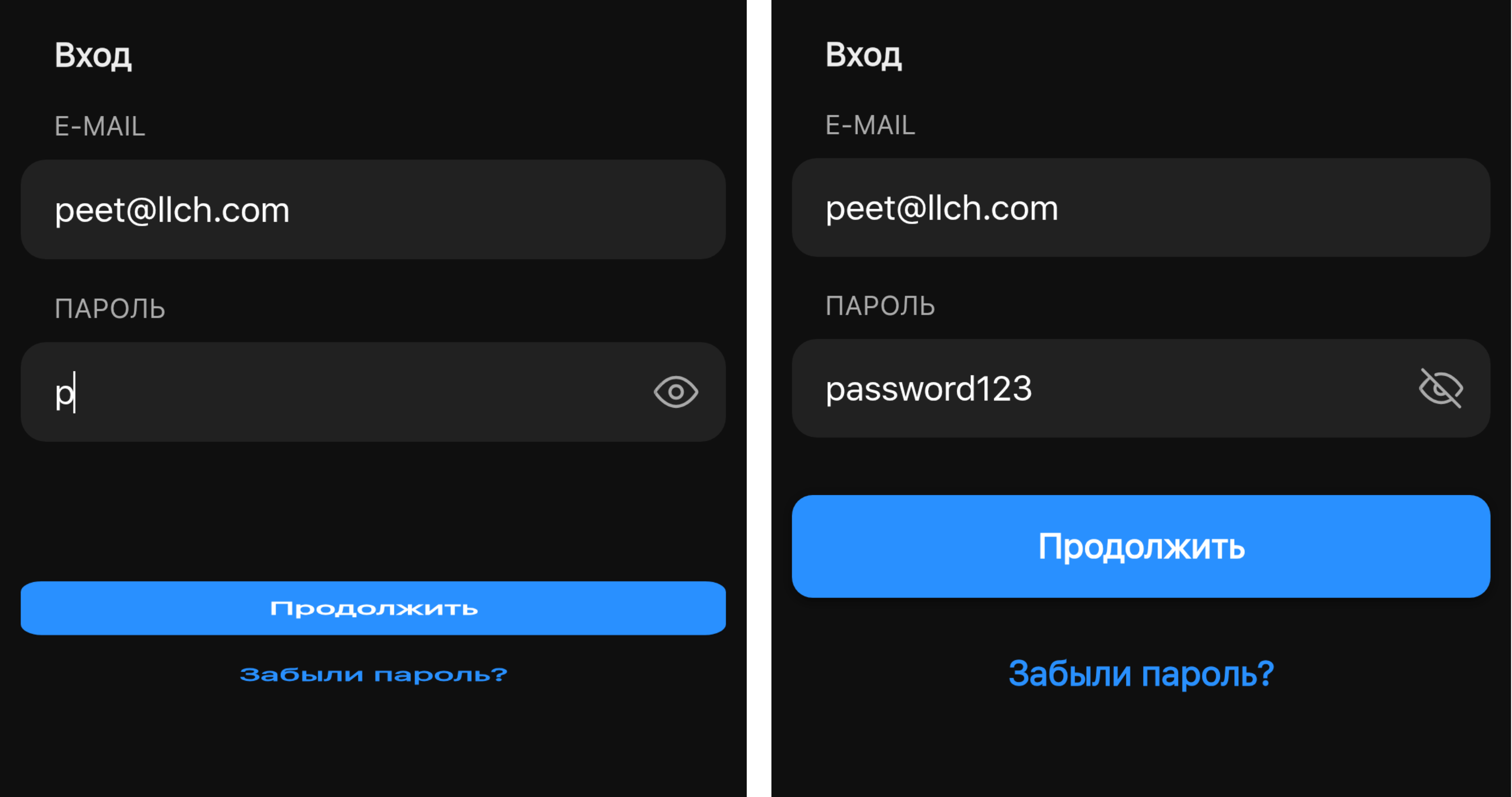}
  \caption{\textbf{Mini-app Screenshot Taken as MAX.} Screenshots taken by invoking Android's screenshot API from within MAX, showing a mini-app login form.}
  \label{fig:password-screenshot}
\end{figure}

The second capture path uses Android’s PixelCopy API to copy the pixels currently displayed in the host's window into an image buffer created by the host's code. It copies from the most recently queued buffer in the Window or Surface behind the displayed UI~\cite{androidpixelcopy}. Even when the WebView isolates web content in a separate renderer process, the GPU services that produce the final pixel output still run inside the host app's process. Thus, renderer-process isolation does not make the rendered pixels opaque to the super-app host.

During our experiments, we found that PixelCopy was more robust to hardware-accelerated rendering state. It returned mini-app UI in 31 of 35 attempts, for an 88.5\% success rate. Crucially, in cases where \texttt{View.draw(Canvas)} returned a blank Bitmap, PixelCopy retrieved valid rendered content, indicating that the displayed surface buffer was available even when the View-level Canvas path failed.

Notably, the UI capture provides information that no other channel can decisively reveal. For example, text entered in a form before submission, auto-complete drop-down selections, and any UI state that exists only in the WebView's visual output. Figure \ref{fig:password-screenshot} shows test credentials entered into a mini-app login form, captured by our instrumentation mid-keystroke. 

Repeatedly capturing screenshots when a user is typing their password will reveal the value even when this field is masked. Similarly, during our tests, we were able to successfully capture frames that showed search queries as they are being typed, previous booking history, and SSO provider selection options---all pre-submission and visible only in the renderer's visual output. At this moment, we did not find evidence that MAX itself invokes these UI-capture paths during normal mini-app use. However, the architecture leaves the mini-app UIs exposed to MAX by default, and if MAX activates this UI capture users and mini-apps would not find out.

\keyinsight{A malicious super-app has a reliable, readily exercisable path to capture mini-app UI, including transient user-visible state.}

\subsection{Access to Mini-app Local Storage}

In Android, WebView persists its state inside the super-app's private data directory, including cookies, localStorage, sessionStorage, and IndexedDB. The super-app controls the WebView container and can interact with it through
WebView management APIs like \texttt{CookieManager} and \texttt{WebStorage}. 

Given this architectural capability, we ask how MAX can access and modify mini-app local storage and whether there are any security barriers. In MAX implementation, when a mini-app wants to persist any data it uses one of these two mechanisms: (1) direct storage, and (2) bridge-mediated storage, where the mini-app can request MAX to store things in its DeviceStorage or SecureStorage.

\myparagraph{Direct Storage}
MAX owns the private directory in which its WebView-based mini-apps store their
cookies and origin-scoped state such as localStorage, sessionStorage, and
IndexedDB. This means that the super-app MAX can access all mini-apps' local storage because it is part of its own. There is no Android security barrier that will prevent MAX from accessing or modifying mini-app local storage.

\begin{figure}[!t]
  \centering
  \includegraphics[width=\columnwidth]{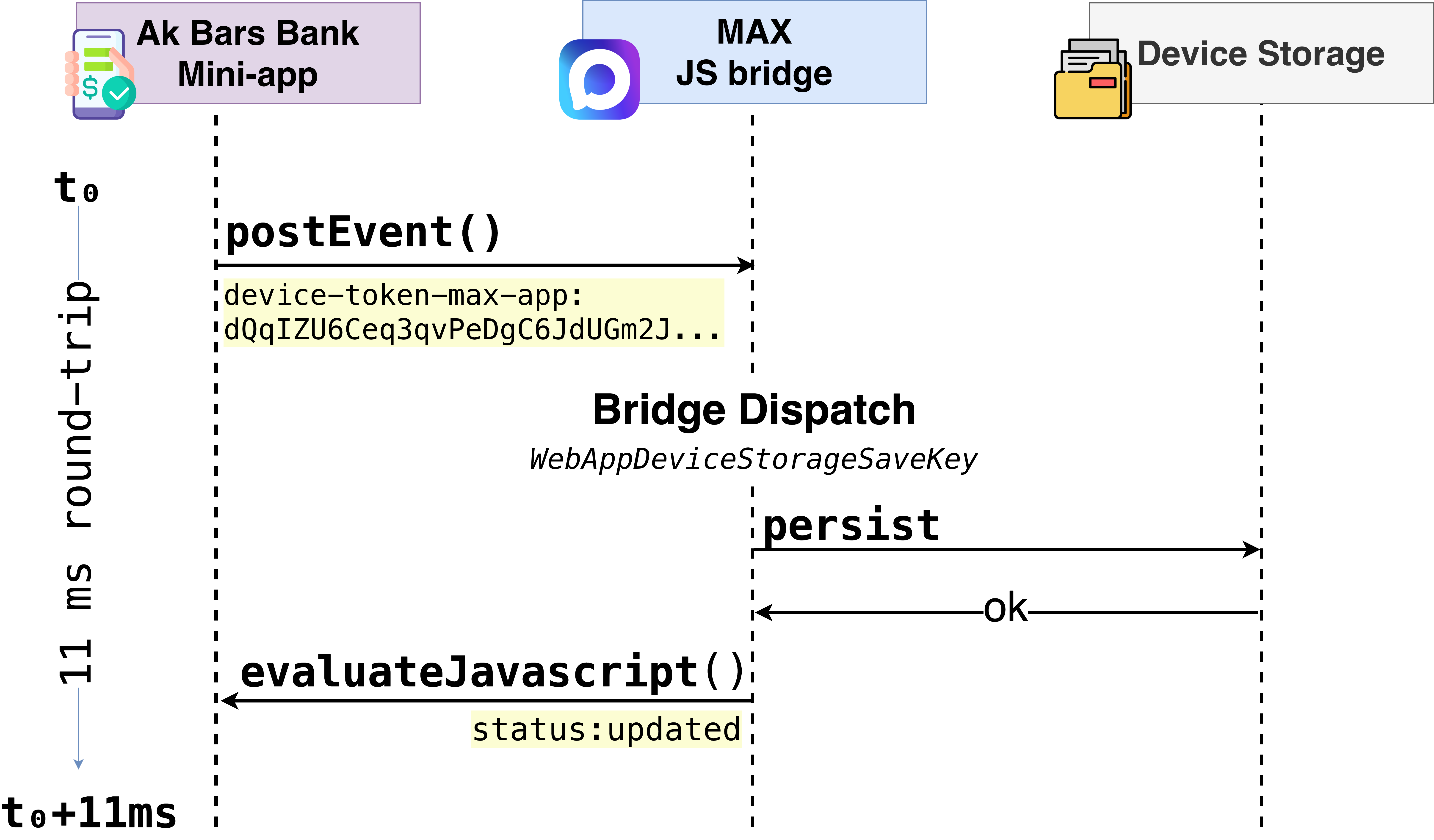}
  \caption{\textbf{DeviceStorage Round-trip.} Ak Bars Bank miniapp persisting a push token via MAX's bridge. MAX receives the token in clear text, persists it, and confirms to the mini-app.}
  \label{fig:device-storage-roundtrip}
\end{figure}
\myparagraph{Bridge-mediated Storage}
MAX does not technically require mini-apps to use the bridge for client-side storage. It enables DOM storage through \texttt{setDomStorageEnabled(true)} in \texttt{WebAppRootScreen}, so mini-apps could store data directly in \texttt{localStorage} or using Web Storage APIs. 

The bridge-mediated storage follows a consistent request–response pattern through the \texttt{WebViewHandler} bridge. First, the mini-app invokes \texttt{postEvent} with a plaintext JSON payload containing the key name and a \texttt{requestId}. Next, MAX's dispatcher receives the request, performs the storage operation on the mini-app's behalf, and constructs a response. Finally, MAX injects the result back into the mini-app via \texttt{evaluateJavascript} invoking \texttt{WebApp.sendEvent} with either the stored value or a structured error.

We found two bridge storage variants implemented:

\textbf{1. Device Storage.} Figure \ref{fig:device-storage-roundtrip} illustrates a typical DeviceStorage round-trip by zooming into the flow of Ak Bars Bank mini-app (\texttt{online.akbars.ru}) persisting a device push token through MAX's bridge. The banking mini-app's 38-byte push token passes through the bridge in plaintext where it persists the value in local storage, and confirms back. Four seconds later, when the mini-app transitions to a new page, a second \texttt{GetKey} for the same token succeeds, completing the mediation cycle. As we can see, every byte of this token is visible to MAX as it brokers the operation.

\begin{figure}[!t]
  \centering
  \includegraphics[width=\columnwidth]{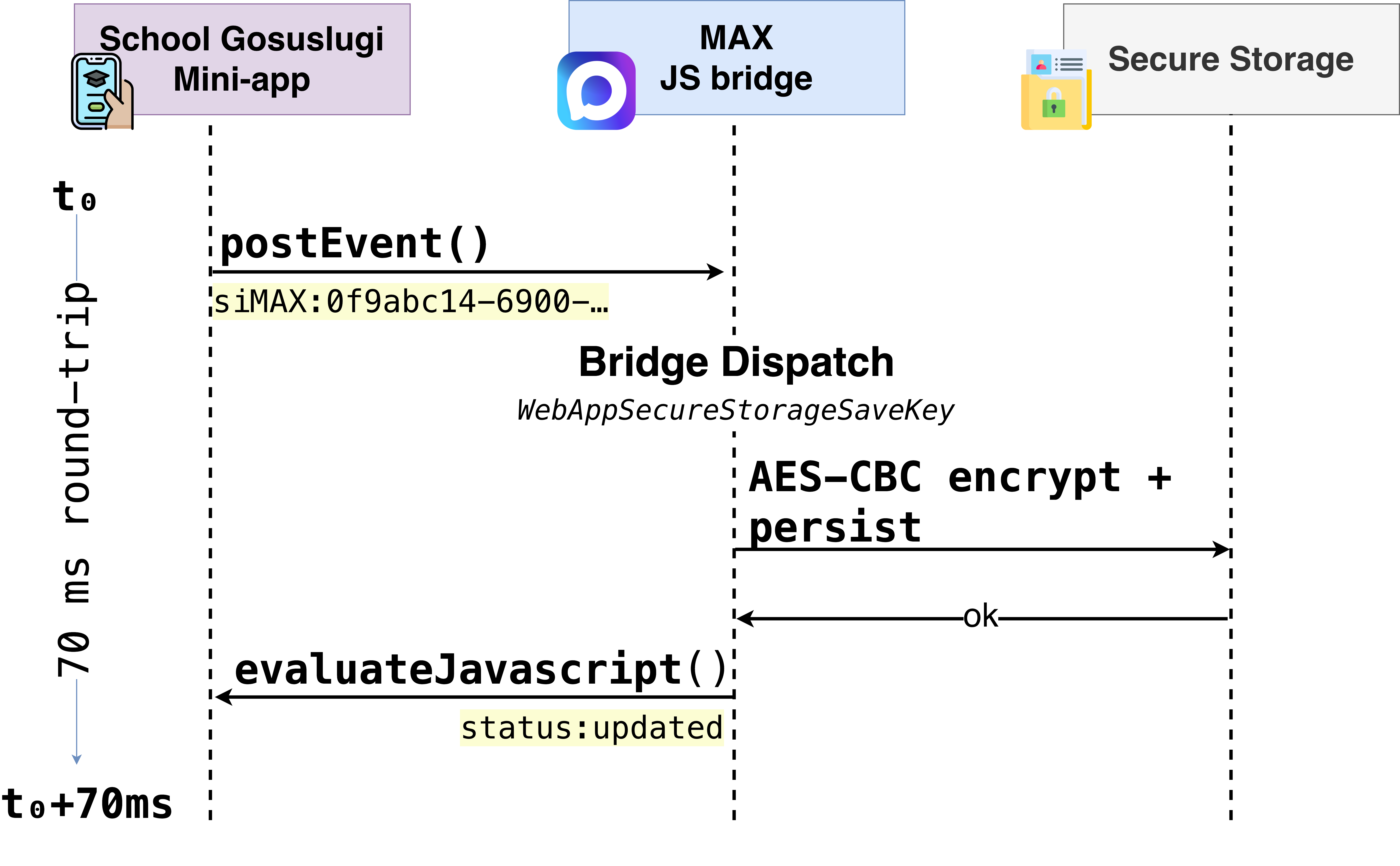}
  \caption{\textbf{SecureStorage Round-trip.} The government school portal persisting a post-ESIA session identifier (\texttt{siMAX} and its UUID value) through MAX's bridge. MAX receives the value, encrypts and stores it into Android's secure storage, and confirms back to the mini-app.}
  \label{fig:secure-storage-roundtrip}
\end{figure}

\textbf{2. Secure Storage} The federal school portal (\code{school-app.gosuslugi.ru}) shows how SecureStorage works in practice (Figure \ref{fig:secure-storage-roundtrip}). During Unified System of Identification and Authentication (ESIA) login, the mini-app stores a session key \texttt{siMAX} through MAX’s bridge. Although the final on-disk value is protected using Android Keystore-backed encryption, the ESIA session key first crosses the host bridge in plaintext, where MAX observes it before encrypting and persisting it.

The pattern is consistent across sectors. The federal culture portal stores a device UUID, an API key, and a geocoder API key through \texttt{DeviceStorage}, and stores a per-user auth flag and a 908-byte telemetry batch through \texttt{SecureStorage}. Nine healthcare mini-apps request the user's MAX authentication token through \texttt{DeviceStorage}. In every read and write, regardless of storage variant, the value transits through MAX in plaintext.

Despite the fact that a mini-app can persist data directly, every mini-app we tested used MAX’s bridge API for persistence rather than the private DOM storage. This turns MAX into a plaintext storage broker that sees every key--value pair mini-apps store, retrieve, or delete.

\keyinsight{A malicious super-app has several ways to read and write any information that a mini-app wants to persist on device.}

\subsection{JavaScript Injection Into Mini-app}
In Android, \texttt{evaluateJavascript} exposes a direct code-injection channel from the super-app into a mini-app's currently loaded page. The API evaluates JavaScript in the context of the currently displayed page, making it such that the super-app-supplied code executes with the same page-level authority as the mini-app's own scripts~\cite{webview}. It can access the DOM, window, global variables, and any Web APIs available to the page context.

Given this architectural capability, we ask how MAX can inject its own code into a mini-app's runtime. In MAX, the communication channel between mini-app and super-app is bidirectional but asymmetric. In the mini-app to super-app direction, the surface is narrow and fully fixed: two JS bridges expose three callable methods: \texttt{postEvent} and \texttt{resolveShare} on \texttt{WebViewHandler}, and \texttt{trackFcp} on \texttt{AndroidPerf}. These are the only entry points through which mini-app JavaScript can reach the super-app's Java process. Both of these bridges are torn down on mini-app exit via \texttt{removeJavascriptInterface}. In the super-app to mini-app direction, the surface is unbounded. This means, \texttt{evaluateJavascript} can execute MAX's code in the mini-app's context, with no restriction on content, timing, or frequency. During our experiments, we observed MAX use this capability for three purposes.

\begin{figure}[!t]
  \centering
  \includegraphics[width=\columnwidth]{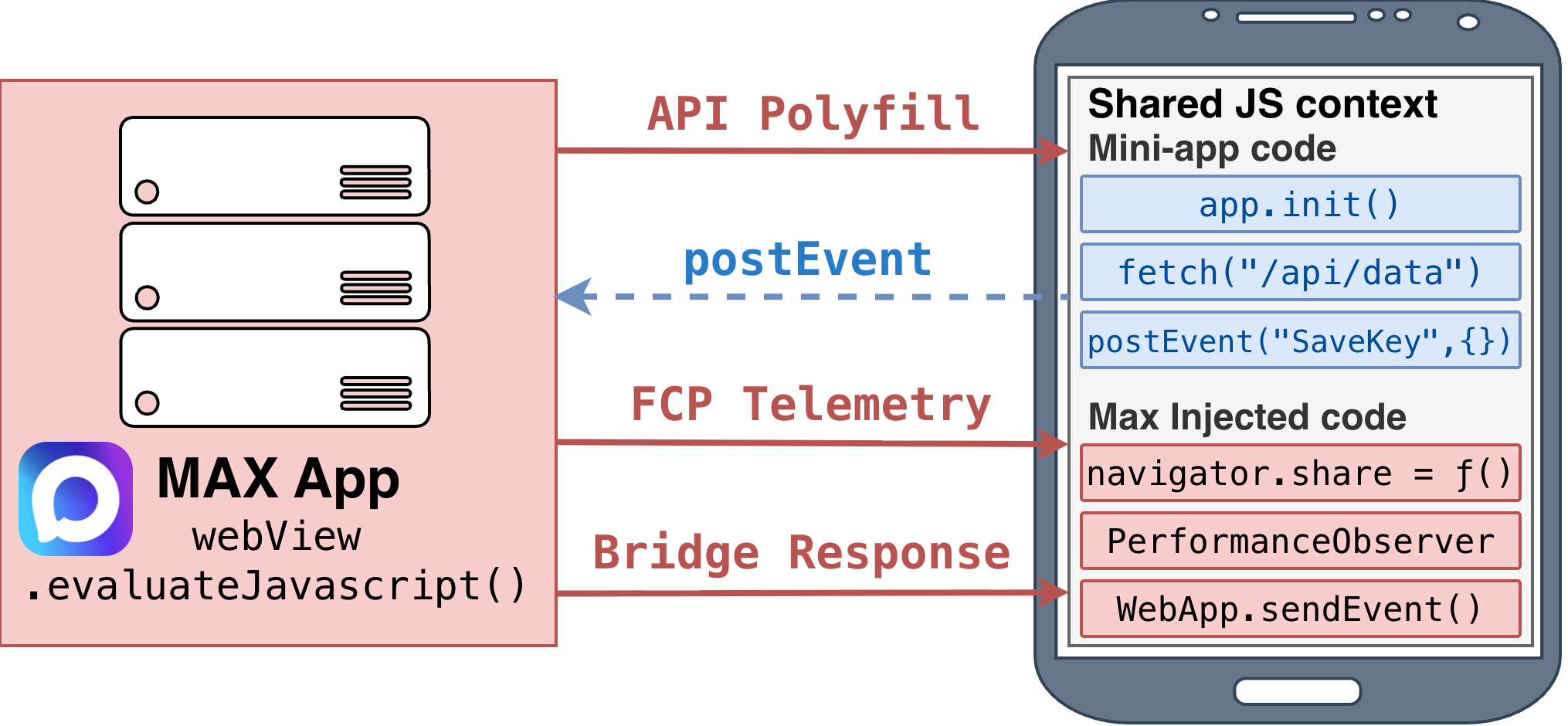}
  \caption{\textbf{JavaScript Injection.} MAX injects three categories of JavaScript into the mini-app's WebView via \texttt{evaluateJavascript}: an API polyfill that reroutes file-sharing through MAX, a performance probe that reports page load timing back to MAX, and bridge responses.}
  \label{fig:js-injection}
\end{figure}

\myparagraph{API Polyfilling}
Mini-apps call the Web Share API, \texttt{navigator.share}, to allow users to share documents, images, tickets, links, or other
mini-app content through Android's native sharing interface. Since Android WebView does not provide this API, we observed that MAX injects a compatibility polyfill into hosted mini-app pages. Unfortunately, this replacement is invisible to the mini-app because the standard API signature is preserved by the polyfill.

When a mini-app shares a file using this method, the polyfill reads the file content and routes it through the bridge. MAX receives the full file bytes inline before any sharing occurs. Files exceeding 3 MiB are silently dropped with no error callback. MAX arms the Web Share polyfill on every page-load, but we did not observe the function being invoked by any mini-apps during our captures.

\myparagraph{Performance Telemetry}
\label{subsec:perftelemetry}
When a mini-app loads a new page, we observed that MAX injects a one-shot probe that measures \texttt{first-contentful-paint} and reports the result back to it via a dedicated JavaScript bridge. The probe is not injected uniformly for all mini-apps, suggesting host or session-level configuration. This is used to capture the performance of each page by observing the time (in milliseconds) since navigation start.

\myparagraph{Bridge Response}
MAX sends data to a mini-app by injecting JavaScript into the mini-app's context. Most bridge calls follow a normal API request-response flow: the mini-app calls \texttt{postEvent}, and MAX returns the response through \texttt{WebApp.sendEvent} with a matching \texttt{requestId}. For others, MAX unilaterally sends data to the mini-app: For example, we observed \texttt{WebAppBackButtonPressed} event was MAX-initiated. The mini-app receives both bridge responses and super-app initiated events through the same mechanism. Unless the mini-app implements its own provenance checks, it has no built-in mechanism to verify whether an incoming event is a genuine reply to a request it made or super-app initiated event. 

The response channel also carries identity material. We observed three distinct identifiers in transit through bridge responses: \texttt{device-token-max-app}: Persistent per-device token, \texttt{siMAX}: post-ESIA session UUID, and \texttt{delegatedAuthorizationCounter}: ESIA auth-state counter.
\begingroup
\rowcolors{2}{gray!10}{white}
\begin{table}[!t]
\centering
\renewcommand{\arraystretch}{1.1}
\setlength{\tabcolsep}{4pt}
\scriptsize
\begin{tabular}{@{}lp{0.45\columnwidth}@{}}
\toprule
\textbf{Event} & \textbf{Purpose} \\
\midrule
\texttt{DeviceStorageGetKey}        & key--value read response \\
\texttt{DeviceStorageSaveKey}       & key--value write ack \\
\texttt{SecureStorageGetKey}        & Encrypted key--value read response \\
\texttt{SecureStorageSaveKey}       & Encrypted key--value write ack \\
\texttt{BackButtonPressed}          & Synthetic UI event \\
\texttt{HapticFeedbackImpact}       & Haptic impact ack \\
\texttt{HapticFeedbackNotification} & Haptic notification ack \\
\texttt{MaxShare}                   & Share completion \\
\texttt{DownloadFile}               & Download state \\
\texttt{RequestPhone}               & Phone-number request response \\
\bottomrule
\end{tabular}
\caption{\textbf{Bridge Response Events.} All event names share the prefix \texttt{WebApp}.}
\label{tab:bridge-response-events}
\end{table}
\endgroup

Overall, while this JS injection capability currently provides legitimate functionality to mini-apps, it gives the super-app control over both code and data flowing into the mini-app. First, MAX can run JavaScript in the mini-app's context, with full access to the DOM, window, and all Web APIs. The polyfill and telemetry injections documented above confirm that MAX is actively exercising this capability. Next, super-app injected events arrive through the same \texttt{WebApp.sendEvent} interface as responses to mini-app requests, making them indistinguishable from a genuine reply. Lastly, the response channel carries sensitive material like persistent device identifiers, storage values, and authentication state in plaintext. The mini-app must trust that these values are authentic because it has no mechanism to verify them independently.

\keyinsight{A malicious super-app can silently inject and execute arbitrary JavaScript inside a mini-app, with no mechanism for the mini-app to distinguish injected code or data.}

\subsection{Interception of Mini-app Network Activity}

In Android, a mini-app's HTTP requests do not traverse the super-app's HTTP client, connection pool, or socket layer and should not appear in the super-app's process-level network statistics. Instrumenting only the super-app's network surface would lead one to conclude that the mini-app traffic is completely invisible to the super-app. However, the WebView client APIs, including \texttt{WebViewClient}, \texttt{WebChromeClient}, and the \texttt{@JavascriptInterface} bridges registered, run in the super-app process by design.

Given this architectural capability, we ask whether MAX can use it to intercept a mini-app's network activities. In MAX, the aforementioned APIs can be invoked for every observable event in a mini-app's lifecycle. Taken together, the callbacks in response to WebView lifecycle, navigation, resource, console, and error events reconstruct an almost complete activity trace of every mini-app session. 
During our experiments we observed several observational channels (see Table \ref{tab:network-calls}):
\begin{figure}[!t]
  \centering
  \includegraphics[width=\columnwidth]{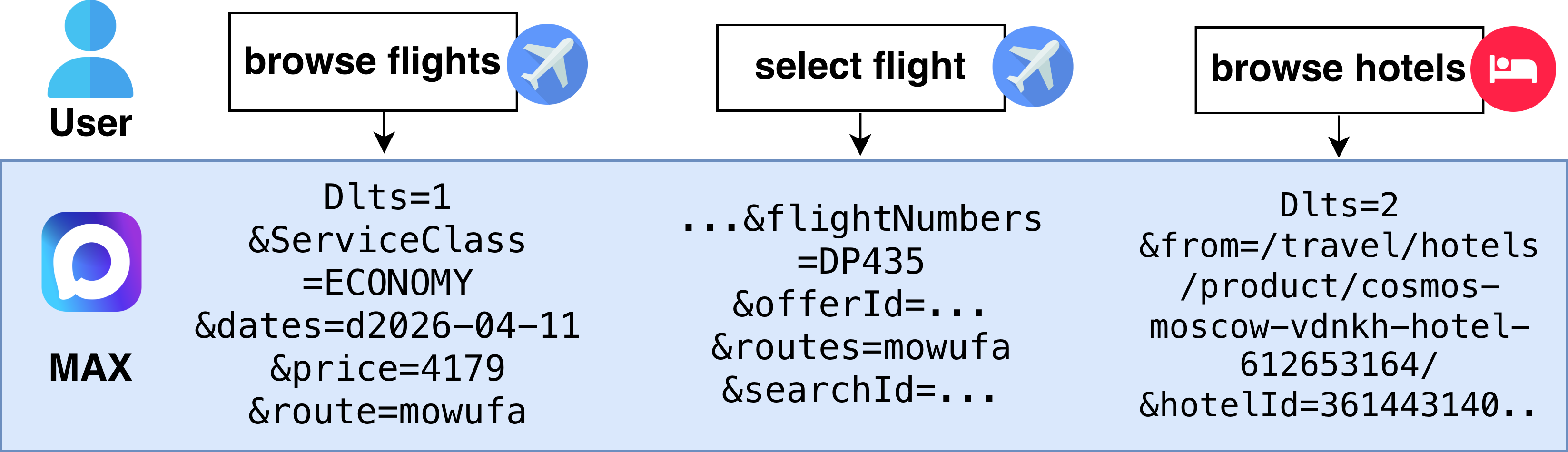}
  \caption{\textbf{Network Interception.} Intercepted mini-app requests reveal important travel booking details like origin/destination, dates, party composition, the specific Pobeda flight DP435 with offerId, and a named hotel property.}
  \label{fig:should-intercept-req}
\end{figure}

\textbf{(1) Resource Requests.} MAX observes each mini-app's requests for resources before they leave the WebView and has the ability to block, rewrite, or inject content into any of these requests.

\textbf{(2) URL Loads.} On every page navigation initiated by the webpage, user, or an HTTP redirect, MAX is presented with the full destination URL. This includes the mini-app’s initialization payload containing user ID, chat ID, authentication date, and an HMAC tag passed as a URL fragment.

\textbf{(3) Network \& HTTP Errors.} MAX gets a detailed view of its mini-app's operational health as it reviews private error information via callbacks. This includes HTTP errors, host name resolution failures, SSL protocol failures, connection errors, and time-outs.

\textbf{(4) Page Lifecycle \& Performance.} MAX receives a timestamped log of every page the user visits within a mini-app, at millisecond resolution. Combined with the performance telemetry (Section \ref{subsec:perftelemetry}), MAX has complete details about the entire lifecycle and performance of each page in a mini-app.

\textbf{(5) Application Runtime Outputs.} MAX receives console logs from the JavaScript calls made by its mini-apps. Through our instrumentation, we observed several critical details leaking from these console log messages, including the user ID, authentication errors, and, in some cases, the full authentication payload.

\textbf{(6) File Download.} When a mini-app starts a file download via the bridge, the file bytes transit through MAX’s process rather than being delivered directly to the mini-app.

Altogether, these methods give MAX a comprehensive runtime view of every mini-app session: which pages the user visits, what resources they fetch, which API endpoints
they contact, what errors occur, what the developer logged to the console, and what files the user downloads. The WebView’s process isolation protects the mini-app’s JavaScript heap and DOM from direct super-app inspection but nearly everything the mini-app does is reported to the super-app through the WebView client API surface.

\keyinsight{A malicious super-app can intercept a mini-app's network requests and reconstruct a detailed image of its activities at runtime, including page loads, resource requests, errors, console logs, and downloads.}

\subsection{Impersonation of User to Mini-app}
In a standard Android app, the app-server authenticates the user directly through its own login flow, its own session tokens, and direct integration with identity providers. Typically, an app does not rely on its hosting environment to vouch for who the user is. However, a super-app interposes itself between the user and the mini-app, issuing identity on the user's behalf and mediating access to device capabilities through a bridge API.

Given this architectural capability, we ask whether MAX has identity attributes that can enable it to impersonate a user to mini-apps.
During our experiments, we observed three points where MAX sits between a user and a mini-app for authentication and authorization: (1) super-app issued identification, (2) third-party authentication, (3) user consent.

\myparagraph{Super-app Issued Identification}
Every mini-app page load arrives with a \texttt{WebAppData} URL fragment that MAX constructs and appends to the mini-app's URL. The bundle contains six identity fields: \texttt{user.id}, \texttt{chat.id}, \texttt{auth\_date}, \texttt{query\_id}, \texttt{ip}, and an HMAC integrity token as a \texttt{hash}. Beyond the per-session \texttt{initData}, MAX issues persistent identifiers that mini-apps store and retrieve through the bridge. 

\myparagraph{Third-party Authentication}
When a mini-app authenticates via an external provider, MAX can observe authentication material that should remain between the mini-app and the identity provider. We observe this at multiple layers: mini-apps pass authentication URLs through MAX before opening them in the system browser, share WebView cookies across unrelated mini-app contexts, expose credentials in console output captured by MAX, and store cookies and web storage under MAX’s UID. An illustrative case that we saw in our captures was the federal school portal (\texttt{school-app.gosuslugi.ru}) initiating ESIA authentication through \texttt{WebAppOpenLink} with a URL containing a \texttt{client\_secret} parameter that contained approximately 4 KB of X.509 certificate material from Russia's Roskazna PKI. As a result, we saw the full authentication bundle for the state services portal, including the cryptographic client credential transit through MAX before the system browser opened.

\begin{figure}[!t]
\centering
\includegraphics[width=\columnwidth]{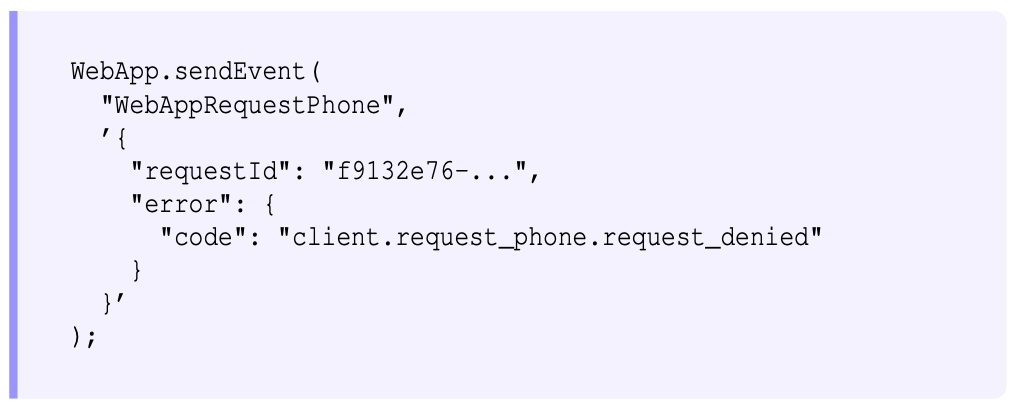}
\caption{\textbf{MAX's Consent Response.} MAX's response to a mini-app when the user declines the phone-share consent dialog.}
\label{fig:webapp-request-phone-deny}
\end{figure}

\myparagraph{User Consent}
All mini-app permission requests are mediated by MAX. Mini-apps request for permissions via the JS bridge, while MAX holds the underlying permission and returns its response through the \texttt{evaluateJavascript} channel. As a result, the mini-app cannot independently verify whether the returned decision matches the user’s choice. For example, \texttt{WebAppRequestPhone} allows a mini-app to request the user’s phone number; MAX is responsible for displaying the consent dialog and returning the response to the mini-app.

MAX sees the following identity attributes during mini-app interactions: the user's \texttt{user.id} and persistent per-device tokens issued by MAX itself, and JWTs, OAuth tokens, ESIA credentials, or session cookies the mini-app accumulated, visible in transit, readable from the shared cookie jar, or stored in bridge-mediated storage. The mini-app does not seem to have any independent verification path for the user's identity.

\keyinsight{A malicious super-app has everything it needs to present itself as the user to any hosted mini-app, or to present a fabricated response as though it came from the mini-app's own backend. This opens up a channel that can be misused by a super-app with malicious intent to impersonate a user to a mini-app.}

\section{Super-apps within Android's Threat Model }
\label{sec:androidvssuperapps}

In this section, we use Mayrhofer et al.'s account of Android's security model as a baseline for analyzing super-apps \cite{androidsecurityplatform}. Their model recognized that Android must give developers enough flexibility to support diverse application patterns, while also providing guardrails that preserve interoperability and prevent misuse of OS mechanisms. 
Super-apps sit exactly at this intersection, where the design provides significant convenience but lacks meaningful guardrails to prevent architectural misuse.

Android's security relies on its ability to provide adequate isolation between untrusted applications. 
A core assumption is that each app is distributed as a \textit{single} Android package, assigned its own UID by developer, and can therefore be isolated from other apps at the process, storage, and network boundaries. This allows Android, via a combination of runtime permissions, SELinux MAC policies, and standard unix-based filesystem controls, to identify the app requesting a resource, check the permissions granted to the app, and grant or deny access.

Since the OS can no longer mediate permissions or provide isolation, the super-app architecture weakens Android's ability to provide any guarantees to the mini-app. 
A super-app is \textit{still} isolated from other Android apps installed on the device, but Android cannot mediate mini-apps as separate apps. Though Android's security model anticipates ``external web resources loaded by an embedded WebView,''  it treats them as belonging to the hosting app's domain~\cite{androidsecurityplatform}. Mini-apps, however, are not webpages: they can implement complicated app-like functionality through super-app mediated APIs.

As we discussed in Section \ref{tab:results}, the super-app has minimal isolation from the mini-apps, allowing for several active attacks. It can screenshot mini-app content, read and write all of its local storage, inject arbitrary JavaScript, mediate all network traffic, and control its auth tokens to silently impersonate any user to any mini-app. 

Another natural consequence of this isolation collapse is the loss of \textit{multi-party authorization}. 
Android's security model is based on multi-party authorization which means an action should only be permitted if all parties (user, developer, and platform) authorize it~\cite{androidsecurityplatform}. When a native app requests for a permission on Android, the user and the platform may veto granting that permission, and the platform needs to enforce that decision. However, when a mini-app asks for a permission, the platform and the user have no visibility or control over granting or denying that permission, and the super-app remains the sole arbiter.

Further, Android's open-ecosystem assumes that third-party code remains ``packaged, signed, and reviewed before execution''~\cite{androidsecurityplatform}. In contrast, a super-app is capable of dynamically loading mini-app code at runtime, and Android has no way to attest what code actually runs. Viewed through the security model's own threat assumptions, the super-app is positioned to mount the very threats against its mini-apps that the Android tries to neutralize between two apps. 

This raises a crucial question: What is Android's position in the super-app architecture? Android remains the platform for the host app, but it is no longer serving as a platform for the mini-app. For mini-apps, the super-app becomes the trusted computing base providing runtime, storage, network, and permission mediation. This creates an authorization collapse. Android can mediate the super-app's access to the device capabilities, but it cannot directly mediate a mini-app's access to these capabilities. From Android's perspective, all the permissions requested by the super-app are reasonable given its delegated role as a platform. It is equally impossible to statically audit if the super-app manages these permissions appropriately, given the mini-apps that may use it will be dynamically loaded at run time. Finally, this trusted computing base is unverified by Android and unconstrained by any compatibility requirements, platform integrity checks, and system-level access control rules. It remains an ordinary third-party app, yet mini-apps must trust it with OS-like authority.

\section{Recommendations}
\label{sec:recommendations}
In this section, we propose a set of defenses aimed at constraining the super-app's unchecked capabilities. Our results show how malicious super-apps have dangerous \mbox{capabilities} that silently undermine the security and privacy of mini-apps and users. They also show how these capabilities violate the core security assumptions of the Android security model. 

We offer near-term strategies as well as longer-term architectural responses and discuss their relative strengths and weaknesses. Our recommendations are designed around the premise that a super-app enforcing its own guardrails is insufficient specifically when a super-app operates under government pressure. Meaningful protections must therefore come from the operating system, or the app store.

\noindent \textbf{Fine-grained Permissions.} 
Android could offer immediate relief by allowing users to revoke \texttt{INTERNET} and \texttt{ACCESS\_NETWORK\_STATE} permissions, or ``allowed only while in use'', as we currently have for fine-grained location. Paired with a restriction on background execution for such apps would prevent a disabled super-app from passively collecting data while offline and exfiltrating it once connectivity is restored. 

\noindent \textbf{Architectural Changes.} 
In the longer term, the OS can close the visibility and isolation gaps it currently leaves unaddressed. We propose three escalating classes of defense: transparency, auditability, and architectural redesign.

First, we recommend user-facing transparency mechanisms. The OS should notify the user when the super-app captures a mini-app's UI. This also naturally extends naturally to two further disclosures that are omitted by the current model. First is an explicit, OS-level indication when a user enters a super-app mediated context, similar to the security indicators indicating VPN or location use. Ideally, there should be an OS-level list of all installed mini-apps, so that the user has the ability to see and revoke permissions through an OS interface. 

Second, the OS should enable auditability through OS-enforced mechanisms that detects when a super-app accesses or modifies mini-app content, storage, or traffic. For example, every time \texttt{shouldInterceptRequest} is called, the function should be logged along with the resource request intercepted by the super-app. Logs of super-app to mini-app actions should be tamper-resistant and written such that when the super-app misbehaves, it leaves evidence the OS can attest.

Third, we call for innovation in the super-app architecture that lets mini-apps establish trust independently of the super-app. Structurally, the OS should create delineation and restriction in the way it imposes security barriers among apps. With that delineation, the OS should provide per-mini-app sandboxing similar to browser site isolation, assigning each mini-app a distinct security principal with its own storage partition and process boundary so that the super-app can no longer read across mini-apps. It should offer OS-mediated secure channels through which a mini-app reaches its own backend without any network intermediaries, so that even a malicious super-app cannot read or alter the traffic.

Finally, there should be some attestation schemes through which a mini-app's backend can verify it is talking to genuine, unmodified mini-app code, and signed mini-app bundles recorded in a transparency log, restoring the provenance and tamper-evidence guarantees that APK signing provides today.

\section{Conclusion}
In this paper, we analyze the capabilities of a super-app architecture using a case study of Russia's state-backed super-app. Through detailed instrumentation and reverse engineering, we show dangerous capabilities available to a super-app to silently act as a man-in-the-middle for all the mini-app interactions. While we didn't observe MAX actively exploiting these capabilities maliciously, none the less our finding reveal that MAX, like any other super-app, is architecturally capable of exercising them. 

With all mini-apps running under a single Android principal, Android security model does not have any built-in mechanisms to constrain or audit a super-app. Users and mini-app developers are similarly blind, and do not have a way to meaningfully defend against a malicious super-app. We therefore argue that super-apps should not be treated as ordinary apps by mobile operating systems and app stores. 

The mobile OS needs defenses that make super-app's mediation of mini-apps constrained and easily auditable. In the near term, Android should give users stronger control over internet access and background execution. Long term, the mobile OS should support increased transparency, tamper-resistant audit logs, and architectural isolation between the super-app and mini-apps.

%-------------------------------------------------------------------------------
%\section*{Acknowledgments}
%-------------------------------------------------------------------------------

%-------------------------------------------------------------------------------
% optional clearing of the page
\cleardoublepage
\appendix

\bibliographystyle{plain}
\bibliography{superapp}
\cleardoublepage
\appendix
\section{Appendix}
\section*{Ethics Considerations}

There are several aspects of this study where we considered the ethical risks that could arise from our work. The goal of this study is to analyze the super-app architecture, zooming into the security risks it poses for its users. Therefore, we designed and conducted this study following the ethical principles of the Menlo Report~\cite{2012-dittrich-mraf} and in accordance with community norms established by similar measurement work~\cite{Xue2022b,Ramesh2023a,Narayanan2015a,Crandall2015a,Wu2025a}.
Below we identify the stakeholders potentially affected by our work, the harms and benefits relevant to each, the safeguards we adopted, and our justification for conducting and publishing this research.

\paragraph{Stakeholders.} 
We identify four groups of stakeholders: (1) ordinary MAX users in Russia, (2) MAX's operator and the broader MAX platform/ecosystem, (3) the research and activist communities who stand to benefit from an accurate understanding of MAX's region-dependent behavior and VPN-detection practices, (4) the Internet freedom community member who provided our Russian vantage point.

\paragraph{Potential harms and mitigations.}
For our analysis, every interaction with MAX and its mini-apps was generated by test accounts on devices controlled entirely by our research team. We proxied our MAX traffic through a Russian vantage point (VP) to capture MAX's native behavior. The VP was acquired through our collaboration with an Internet freedom community member with years of experience on network traffic analysis in Russia. They have been collaborating with the research community for the last 6 years, and explicitly consented to use their VP for this study. The VP was used solely for the traffic strictly necessary to evaluate region-dependent behavior in MAX. 

We manually interacted with MAX and selected mini-apps from our test accounts as ordinary users and did not run any tests or generate any traffic that would trigger the Russian censorship system.  We also did not probe any Russian infrastructure, or generate large volumes of traffic which could degrade availability for other users. 

\paragraph{Justification}
We believe the safety and privacy risks of this study were minimal by design: no real users were studied, no censorship infrastructure was probed, no traffic volumes large enough to affect availability were generated. Against this limited risk, we weigh the benefit of publicly documenting how a widely used platform could be used for undermining user privacy and security. The generated knowledge is valuable both to the research community studying privacy and surveillance, and ultimately, to the users such mechanisms affect. We therefore believe the balance of risks and benefits favors conducting and publishing this work.

\begingroup
\rowcolors{2}{gray!10}{white}
\begin{table*}[!t]
\centering
\renewcommand{\arraystretch}{1.15}
\setlength{\tabcolsep}{5pt}
\scriptsize
\begin{tabular}{@{}p{0.18\textwidth}p{0.27\textwidth}p{0.50\textwidth}@{}}
\toprule
\textbf{Category} & \textbf{Function} & \textbf{Description} \\
\midrule
Resource Request &
\texttt{shouldInterceptRequest} &
Host inspects a WebView resource request and can optionally supply its own response. \\

URL Load &
\texttt{shouldOverrideUrlLoading} &
Host decides whether WebView should load a navigation URL. \\

Network \& HTTP Errors &
\texttt{onReceivedHttpError} &
Reports HTTP errors, i.e., response status codes $\geq 400$. \\

Network \& HTTP Errors &
\texttt{onReceivedError} &
Reports resource-loading failures, such as connection errors. \\

Network \& HTTP Errors &
\texttt{net::ERR\_CONNECTION\_TIMED\_OUT} &
Chromium error for a timed-out connection attempt. \\

Network \& HTTP Errors &
\texttt{net::ERR\_SSL\_PROTOCOL\_ERROR} &
Chromium error for an SSL protocol failure. \\

Network \& HTTP Errors &
\texttt{net::ERR\_NAME\_NOT\_RESOLVED} &
Chromium error for failed hostname resolution. \\

Page Lifecycle \& Performance &
\texttt{onPageStarted} &
Reports that a load has started. \\

Page Lifecycle \& Performance &
\texttt{onPageFinished} &
Reports that a page load has finished. \\

Application Runtime Outputs &
\texttt{onConsoleMessage} &
Reports JavaScript console messages. \\

File Download &
\texttt{WebAppDownloadFile} &
Bridge function used to support mini-app file downloads. \\

\bottomrule
\end{tabular}
\caption{\textbf{Observed WebView Callbacks and Runtime Events.}}
\label{tab:network-calls}
\end{table*}
\endgroup
\section*{Open Science}
As per the open science policy, we make all of our artifacts available at the following link: \texttt{https://github.com/censoredplanet/rmax}. It contains the sanitized plaintext capture logs that underpin our results, the instrumentation used to collect them, and the decoding and analysis tooling used to process them. A README at the top level details the artifact and folder structure, the software and device requirements, and the format of the capture logs. In the following, we detail all parts of our artifact.

\myparagraph{Instrumentation.} The unified Frida script described in Section 3.2 is located in \texttt{runtime-capture/max-capture-unified-ru.js}. It arms hooks across the four functional layers we describe --- TLS read/write, the JavaScript bridge, WebView callbacks, and sensitive Android APIs --- and detects the app build at startup, selecting the CryptoPro GOST hooks for the Russian build and the Conscrypt/BoringSSL path otherwise. A single script covers both builds. The wrapper \texttt{runtime-capture/max-capture-ru.sh} spawns or attaches to the app, injects the run configuration, and writes a timestamped capture file. 

\myparagraph{Decoding and Analysis.} \texttt{runtime-capture/ingest\_capture.py} implements the decoding and correlation stage of Section 3.4 and the analysis database of Section 3.5. It parses the binary RPC envelope specified in Table 1, joins the \texttt{[EVT]}/\texttt{[DATA]}/\texttt{[PROBE]} triplets by (\texttt{ssl\_id}, \texttt{seq}, \texttt{dir}), reassembles the per-connection byte stream before extracting RPC frames and pairs bridge dispatches with injected responses by \texttt{requestId}, emitting round-trip latencies and retaining unmatched injections as host-initiated events. 

\myparagraph{Capture Logs.} \texttt{runtime-capture/sanitized-data/} contains the sanitized session logs, organized into three capture sets. This is the source data for the capabilities we report in Sections 4.1 through 4.5, and cover the mini-apps listed in Table 4. The log format is documented in detail in the README so the logs can be re-parsed independently of our tooling. 

\myparagraph{Mini-app UI Capture.} \texttt{screenshot-capture/} contains the Frida script used for the UI capture results of Section 4.1 and Figure 5, together with the frames it produced. Every released frame has been manually reviewed.

\myparagraph{Anonymization.} Every capture log has been sanitized to remove material that identifies our test accounts, our devices, or our measurement vantage point: account phone numbers and names, user and session identifiers, authentication tokens and one-time codes, device fingerprints and hardware serials, all IP addresses, and any free-text message content. Values are replaced rather than deleted. Each secret becomes a stable placeholder derived from a keyed hash, so that the same value always yields the same placeholder across the corpus. This preserves the properties our analysis depends on without the underlying value being recoverable. 
\begin{figure}[!t]
\centering
  \includegraphics[width=0.6\columnwidth]{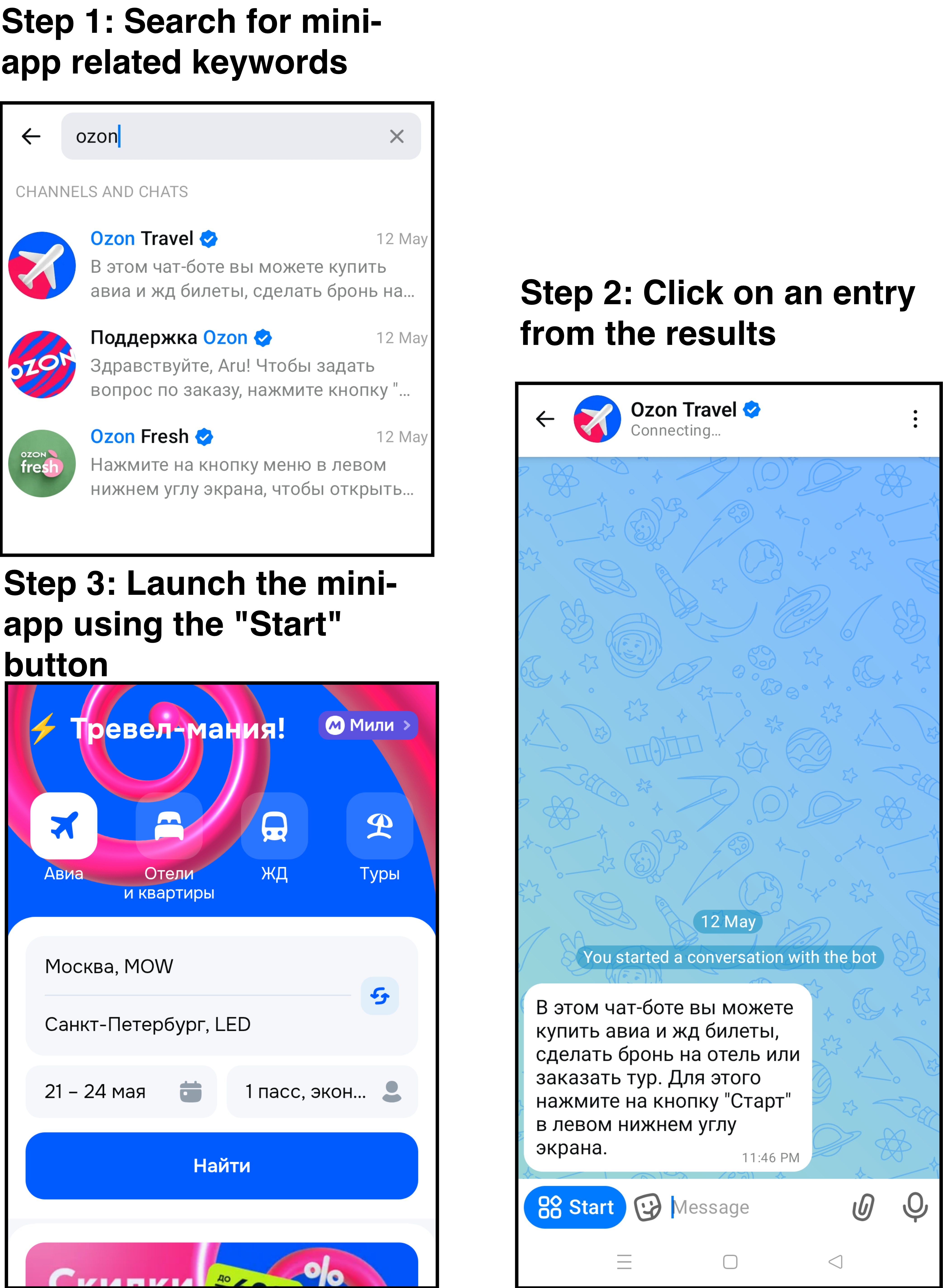}
  \vspace{.5em}
  \caption{\textbf{Mini-app Search.} Navigate to search on Max, and search for the mini-app's name or related keywords; (2) Click on one of the entries from search to launch the associated chatbot; (3) Launch the mini-app from the "Start" button.}
  \label{fig:miniapp-search}
\end{figure}
\onecolumn
\begingroup
\scriptsize
\renewcommand{\arraystretch}{1.25}
\setlength{\tabcolsep}{3pt}

\begin{longtable}{>{\raggedright\arraybackslash}p{0.3\textwidth}>{\raggedright\arraybackslash}p{0.38\textwidth}>{\raggedright\arraybackslash}p{0.2\textwidth}}

\caption{\textbf{Observed MAX Mini-apps.} Names, English names, and bot identifiers are preserved as collected.}
\label{tab:max-miniapp-bots} \\
\toprule
\textbf{Name} & \textbf{English Name} & \textbf{Bot\_id} \\
\midrule
\endfirsthead

\multicolumn{3}{@{}l}{\scriptsize\itshape Continued from previous page} \\
\toprule
\textbf{Name} & \textbf{English Name} & \textbf{Bot\_id} \\
\midrule
\endhead

\midrule
\multicolumn{3}{r@{}}{\scriptsize\itshape Continued on next page} \\
\endfoot

\bottomrule
\endlastfoot

\rowcolor{gray!10}
\makecell[l]{\foreignlanguage{russian}{Национальная электронная}\\\foreignlanguage{russian}{библиотека Республики}\\\foreignlanguage{russian}{Татарстан}} & National Electronic Library of the Republic of Tatarstan & rt\_kitap\_bot \\
\foreignlanguage{russian}{JetBot - конструктор чат-ботов} & JetBot & support\_jet\_bot \\
\rowcolor{gray!10}
Ozon Fresh & Ozon Fresh & ozonfresh\_bot \\
Ozon Travel & Ozon Travel & ozontravel\_bot \\
\rowcolor{gray!10}
\makecell[l]{\foreignlanguage{russian}{Чат-Бот МФЦ Республики}\\\foreignlanguage{russian}{Дагестан}} & Multifunctional Centers (MFC) of the Republic of Dagestan. & mfcrd2025\_bot \\
\makecell[l]{\foreignlanguage{russian}{Чат-бот МФЦ Чеченской}\\\foreignlanguage{russian}{Республики}} & Multifunctional Centers (MFC) of the Republic of Chechen. & mfcchr\_maxbot \\
\rowcolor{gray!10}
\foreignlanguage{russian}{Алиса Al} & Yandex Alice AI & alice \\
\makecell[l]{\foreignlanguage{russian}{Министерство ЖКХ и ТЭ}\\\foreignlanguage{russian}{РСО-Алания}} & Ministry of Housing and Public Utilities, Fuel and Energy of the Republic of North Ossetia-Alania & minjkh\_rso\_a\_bot \\
\rowcolor{gray!10}
Teeeams & Teeeams & teeeams\_bot \\
T2 & T2 & t2russia\_bot \\
\rowcolor{gray!10}
\foreignlanguage{russian}{Сферум} & Sferum & sferum\_bot \\
\foreignlanguage{russian}{Помощник Сферума} & Sferum Assistant & sferum\_help \\
\rowcolor{gray!10}
\makecell[l]{\foreignlanguage{russian}{Мосбилет: афиша}\\\foreignlanguage{russian}{Москвы}} & Mosbilet: Moscow Poster & mosbilet\_bot \\
\foreignlanguage{russian}{Мос.Фото} & Mosphoto & mosfoto\_bot \\
\rowcolor{gray!10}
\foreignlanguage{russian}{ГИС-бот Ямал} & GIS-bot Yamal & gisyamal\_bot \\
\makecell[l]{\foreignlanguage{russian}{Моё здоровье - Запись на}\\\foreignlanguage{russian}{прием - Омская область}} & My Health & kvrachu\_omsk\_bot \\
\rowcolor{gray!10}
\foreignlanguage{russian}{Доктор Миша (запись к врачу в Хабаровском крае)} & Doctor Misha (appointment with a doctor in the Khabarovsk Krai) & mishakhv\_bot \\
\foreignlanguage{russian}{Тыва. Запись к врачу} & Tuva. Doctor's appointment & \makecell[l]{zdrav\_tuva\_bot\\} \\
\rowcolor{gray!10}
\foreignlanguage{russian}{Запись к врачу в Кузбассе} & Appointment with a doctor in Kuzbass & kuzdrav\_vrach42\_bot \\
\foreignlanguage{russian}{Госуслуги Культура} & Gosuslugi Kultura & gosuslugi\_culture\_bot \\
\rowcolor{gray!10}
\foreignlanguage{russian}{Госуслуги Дом} & Gosuslugi Dom & gosuslugi\_dom\_bot \\
\foreignlanguage{russian}{Госуслуги Камчатского края} & Gosuslugi Kamchatka Krai & gosuslugi41\_kamchatka\_bot \\
\rowcolor{gray!10}
\makecell[l]{\foreignlanguage{russian}{Госуслуги Республики}\\\foreignlanguage{russian}{Татарстан}} & State Services Portal of the Republic of Tatarstan & UslugiRT\_bot \\
\foreignlanguage{russian}{Госуслуги. Моя школа} & Gosuslugi Moya Shkola & school\_bot \\
\rowcolor{gray!10}
\foreignlanguage{russian}{Госуслуги Решаем Вместе} & Gosuslugi. We Decide Together & reshaem\_vmeste\_bot \\
\makecell[l]{\foreignlanguage{russian}{«Мое здоровье»}\\\foreignlanguage{russian}{Магаданская область}} & "My Health" Magadan Oblast & magadanzdrav\_bot \\
\rowcolor{gray!10}
\makecell[l]{\foreignlanguage{russian}{Мое здоровье Республика}\\\foreignlanguage{russian}{Алтай}} & My Health Republic of Altai & minzdrav\_respaltay\_bot \\
\foreignlanguage{russian}{Смоленск. Моё здоровье} & Smolensk. My Health & reg67\_medicina\_bot \\
\rowcolor{gray!10}
\foreignlanguage{russian}{Мое здоровье. Забайкалье} & My health. Zabaykalye & \makecell[l]{zapv\_75\_bot\\} \\
RUSSPASS \foreignlanguage{russian}{Бот} & RUSSPASS Bot & russpass\_bot \\
\rowcolor{gray!10}
\makecell[l]{\foreignlanguage{russian}{Моё здоровье.}\\\foreignlanguage{russian}{Ленинградская область}} & My health. Leningrad Region & doctor\_lenobl\_bot \\
\foreignlanguage{russian}{Интервидение 2025} & Intervision 2025 & intervision\_2025\_bot \\
\rowcolor{gray!10}
\makecell[l]{\foreignlanguage{russian}{Сервис «Мое здоровье»}\\\foreignlanguage{russian}{Мурманской области}} & \makecell[l]{"My Health" Service\\Murmansk Region} & mur\_zdrav51\_bot \\
\foreignlanguage{russian}{Электронные сервисы в здравоохранении области Курской} & Electronic services in healthcare of the Kursk region & kurskzdrav\_bot \\
\rowcolor{gray!10}
\foreignlanguage{russian}{Запись к врачу СПб -Служба 122} & Doctor's appointment St. Petersburg - Service 122 & \makecell[l]{spb\_122\_bot\\} \\
\foreignlanguage{russian}{Новогодний адвент-календарь МАХ} & New Year's Advent Calendar MAX & max\_advent\_2025\_bot \\
\rowcolor{gray!10}
\makecell[l]{\foreignlanguage{russian}{Запись к врачу и Телемедицина.}\\\foreignlanguage{russian}{Краснодарский край}} & Doctor's appointment and Telemedicine. Krasnodar Krai & zapis\_na\_priem\_kk\_bot \\
\foreignlanguage{russian}{Бессмертный полк онлайн} & Immortal Regiment Online & polk\_app\_bot \\
\rowcolor{gray!10}
\makecell[l]{\foreignlanguage{russian}{Телемедицинские консультации}\\\foreignlanguage{russian}{Архангельской области}} & Telemedicine consultations of the Arkhangelsk Region & zdrav29\_bot \\
\foreignlanguage{russian}{Город заданий} & City of Tasks & gorod\_zadaniy\_bot \\
\rowcolor{gray!10}
\makecell[l]{\foreignlanguage{russian}{Моё здоровье Ростовская}\\\foreignlanguage{russian}{область}} & My health Rostov region & \makecell[l]{zdrav\_rnd\_bot\\} \\
\makecell[l]{\foreignlanguage{russian}{Помощник Министерства}\\\foreignlanguage{russian}{Имущества |}\\\foreignlanguage{russian}{Хабаровск}} & Ministry of Property Assistance | Khabarovsk & min\_imushestvo\_khabarovsky\_kray\_bot \\
\rowcolor{gray!10}
\foreignlanguage{russian}{ЕКЦ СФР} &  & sfr\_chat\_bot \\
\foreignlanguage{russian}{WiFi в ЯНАО} & Yamalo-Nenets Autonomous Okrug (YNAO) & free\_yanao\_bot \\
\rowcolor{gray!10}
\makecell[l]{\foreignlanguage{russian}{Амбассадоры Пушкинской}\\\foreignlanguage{russian}{карты}} & Ambassadors of the Pushkin Card & ambassador\_pushkincard\_bot \\
\foreignlanguage{russian}{Чат-бот Республики Адыгея "Бэла"} & Republic of Adygea "Bella" & adg\_bela\_bot \\
\rowcolor{gray!10}
\foreignlanguage{russian}{Конструктор открыток} & Post Card Builder & maxpostcards\_bot \\
\foreignlanguage{russian}{Ак Барс Онлайн} & Ak Bars Online & akbarsonline\_bot \\
\rowcolor{gray!10}
\foreignlanguage{russian}{Страховая компания «Согласие»} & Soglasie Insurance Company & soglasie\_bot \\
\foreignlanguage{russian}{Знание. Премия: народное голосование} & Znanie.Award & znanie\_premiya\_bot. \\
\rowcolor{gray!10}
\foreignlanguage{russian}{Переменка} & Peremenka & \makecell[l]{peremenka\_bot\\} \\
\foreignlanguage{russian}{Банк Русский Стандарт} & Russian Standard Bank & bank\_russian\_standart\_bot \\
\rowcolor{gray!10}
\foreignlanguage{russian}{АльфаСтрахование} & AlfaStrakhovanie & alfastrah\_bot \\
\foreignlanguage{russian}{Чат-бот «Мои Документы Москвы».} & My Documents Moscow & mfc\_moscow\_bot \\
\rowcolor{gray!10}
\makecell[l]{\foreignlanguage{russian}{Чат-бот «Активного}\\\foreignlanguage{russian}{гражданина»}} & Active Citizen Chatbot & ag\_mos\_ru\_bot \\
\foreignlanguage{russian}{ЕМИАС.ИНФО} & EMIAS.INFO & \makecell[l]{emiasinfo\_bot\\} \\
\rowcolor{gray!10}
\makecell[l]{\foreignlanguage{russian}{Помощник от}\\\foreignlanguage{russian}{Росгосстраха}} & Rosgosstrakh & \makecell[l]{rosgosstrakh\_bot\\} \\
\foreignlanguage{russian}{Помощник Курского приграничья} & Assistant to the Kursk Borderland & help\_kursk\_bot \\
\rowcolor{gray!10}
\makecell[l]{\foreignlanguage{russian}{Виртуальный помощник по вопросам ЗАГС}\\\foreignlanguage{russian}{Мурманской области}} & Virtual assistant for civil registry office (ZAGS) matters in the Murmansk region & zags\_murman\_bot \\
\foreignlanguage{russian}{Карта жителя РТ} & Citizen Card of the Republic of Tatarstan & citizencardrt\_bot \\
\rowcolor{gray!10}
\makecell[l]{\foreignlanguage{russian}{Мое Здоровье.}\\\foreignlanguage{russian}{Владимирская область}} & My Health. Vladimir Region & k\_vrachu\_33\_bot \\
\foreignlanguage{russian}{Глиф х Дзен} & Dzen & glifaidzen\_bot \\
\rowcolor{gray!10}
\foreignlanguage{russian}{Острова.65 Сахалинская область} & Ostrov.65 Sakhalin Region & sakhalin\_ostrova65\_bot \\
\foreignlanguage{russian}{Мой рецепт Челябинская область} & My Prescription - Chelyabinsk Region & myrecept\_74\_bot \\
\rowcolor{gray!10}
\foreignlanguage{russian}{МФЦ Кировской области} & MFC Kirov Region & mfckirovbot \\
\makecell[l]{\foreignlanguage{russian}{МФЦ Республика}\\\foreignlanguage{russian}{Алтай}} & MFC My Documents & mfcrespaltai\_max\_bot \\
\rowcolor{gray!10}
\foreignlanguage{russian}{Мошенник Не Пройдет} & \makecell[l]{\\The Fraudster Will Not Pass} & stop\_moshennik\_bot \\
\foreignlanguage{russian}{Карта жителя Курской области} & Kursk Region Resident Card & kgko\_kursk\_bot \\
\rowcolor{gray!10}
\foreignlanguage{russian}{Почта Бизнес} & Pochta Biznes & pochta\_business\_bot \\
\foreignlanguage{russian}{Афиша Тверской области} & Tver Oblast Events & tverafisha\_bot \\
\rowcolor{gray!10}
\foreignlanguage{russian}{МФЦ Республики Ингушетия} & MFC of the Republic of Ingushetia & mfc\_ingushetia\_max\_bot \\
\foreignlanguage{russian}{МФЦ РСО-Алания} & MFC RSO-Alania & mfcrsoalania\_max\_bot \\
\rowcolor{gray!10}
\foreignlanguage{russian}{Чат-бот региональных услуг Кировской области} & Chatbot of regional services of the Kirov region & gosuslugi43\_kirov\_bot \\
\foreignlanguage{russian}{план б} & Plan B & bee\_plan\_b\_bot \\
\rowcolor{gray!10}
\foreignlanguage{russian}{Ингосстрах} & Ingosstrakh & ingos\_bot \\
\foreignlanguage{russian}{Пульс образования} & Pulse of Education & sferumpuls\_bot \\
\rowcolor{gray!10}
Salebot \foreignlanguage{russian}{Техническая поддержка.} & Salebot Technical Support & salebot\_support\_bot \\
\foreignlanguage{russian}{Стикеры в МАХ} & Stickers in MAX & stickers \\
\rowcolor{gray!10}
\foreignlanguage{russian}{Сервис «Медицинский личный кабинет Мурманской области»} & Medical Personal Cabinet of the Murmansk Region & mur\_mlk51\_bot \\
\foreignlanguage{russian}{Купер} & Kuper & kuper\_miniapp\_bot \\
\rowcolor{gray!10}
\foreignlanguage{russian}{Почта России} & Pochta Rossii & pochtarus\_bot \\
\foreignlanguage{russian}{Арсенал услуг 71} & Arsenal of services 71 & arsenaluslug\_bot \\
\rowcolor{gray!10}
\foreignlanguage{russian}{Департамент животного мира Республики Марий Эл} & Department of Wildlife Management of the Mari El Republic & depohot12\_bot \\
\foreignlanguage{russian}{Защитники. Под крылом Архангела. Архангельская область} & Archangel of Spetsnaz & defenders\_ao\_bot \\
\rowcolor{gray!10}
\foreignlanguage{russian}{Поддержка для СВОих в HCO} & Support for Our Own in NSO & svoi\_nso\_bot \\
\foreignlanguage{russian}{Глолайм Школьное питание} & GloLime School Meals & glolime\_bot \\
\rowcolor{gray!10}
\foreignlanguage{russian}{Я здесь живу - приложение} & I live here - the app & spb\_mesto\_bot \\
\foreignlanguage{russian}{Белгород. Рядом} & Near Belgorod & belgorod\_app\_bot \\
\rowcolor{gray!10}
\foreignlanguage{russian}{ВСК | Страховой Дом} & VSK Insurance House & vsk\_insurance\_bot \\
\foreignlanguage{russian}{Открытка для учителя} & Teacher Day Card & teacherday\_bot \\

\end{longtable}

\endgroup
\twocolumn

%%%%%%%%%%%%%%%%%%%%%%%%%%%%%%%%%%%%%%%%%%%%%%%%%%%%%%%%%%%%%%%%%%%%%%%%%%%%%%%%
\end{document}